\documentclass{aa}  

\usepackage{graphicx}
\usepackage{txfonts}
\usepackage{multirow}
\usepackage{colortbl}
\usepackage{float}
\usepackage{hyperref}  
\hypersetup{colorlinks=true,linkcolor=[rgb]{1.,0.2,0.2},citecolor=[rgb]{0.1,0.4,1.},filecolor=[rgb]{0.7,0.2,0.2},urlcolor=[rgb]{0.7,0.2,0.2}}

\usepackage{color}
\definecolor{blue}{rgb}{0., 0., 1}
\definecolor{lightblue}{rgb}{0.1,0.4,1.}

\newcommand {\CL}{PSZ1-G311}
\newcommand {\LT}{\texttt{LensTool}}
\newcommand {\T}{Table\,}
\newcommand {\Sec}{Sec.\,}
\newcommand {\Fig}{Fig.\,}
\newcommand {\Eq}{Eq.\,}
\newcommand {\ppxf}{\texttt{Ppxf}}
\newcommand {\sig}{velocity dispersion}
\newcommand {\sn}{$\left<S/N\right>$}

\begin{document} 
\title{A strong lensing model of the galaxy cluster PSZ1 G311.65-18.48}

\author{
G.V.~Pignataro \inst{\ref{difabo}} \fnmsep\thanks{E-mail: \href{mailto:giada.pignataro@studio.unibo.it}{giada.pignataro@studio.unibo.it}} \and
P.~Bergamini \inst{\ref{inafbo}} \and
M.~Meneghetti \inst{\ref{inafbo}} \and
E.~Vanzella \inst{\ref{inafbo}} \and
F.~Calura \inst{\ref{inafbo}} \and
C.~Grillo \inst{\ref{unimi},\ref{darkcc}} \and
P.~Rosati \inst{\ref{unife},\ref{inafbo}} \and
G.~Angora \inst{\ref{unife},\ref{inafna}} \\
G. Brammer \inst{\ref{NBI}} \and
G.~B.~Caminha \inst{\ref{maxplanck}} \and
A.~Mercurio \inst{\ref{inafna}} \and
M.~Nonino \inst{\ref{inafts}} \and
P.~Tozzi \inst{\ref{arcetri}}
}
\institute{
DIFA -- Dipartimento di Fisica e Astronomia, Università di Bologna, via Gobetti 93/2, I-40129 Bologna, Italy \label{difabo} 
\and
INAF -- OAS, Osservatorio di Astrofisica e Scienza dello Spazio di Bologna, via Gobetti 93/3, I-40129 Bologna, Italy \label{inafbo} 
\and
Dipartimento di Fisica, Universit\`a  degli Studi di Milano, via Celoria 16, I-20133 Milano, Italy \label{unimi}
\and
Dark Cosmology Centre, Niels Bohr Institute, University of Copenhagen, Jagtvej 128, DK-2200 Copenhagen, Denmark \label{darkcc}
\and
Dipartimento di Fisica e Scienze della Terra, Universit\`a degli Studi di Ferrara, via Saragat 1, I-44122 Ferrara, Italy \label{unife}
\and
INAF -- Osservatorio Astronomico di Capodimonte, Via Moiariello 16, I-80131 Napoli, Italy \label{inafna}
\and
Cosmic Dawn Center, Niels Bohr Institute, University of Copenhagen, Juliane Maries Vej 30, DK-2100 Copenhagen \O, Denmark\label{NBI} 
\and
Max-Planck-Institut f\"ur Astrophysik, Karl-Schwarzschild-Str. 1, D-85748 Garching, Germany \label{maxplanck}
\and
INAF -- Osservatorio Astronomico di Trieste, via G. B. Tiepolo 11, I-34143, Trieste, Italy \label{inafts}
\and
INAF -- Osservatorio Astrofisico di Arcetri, Largo E. Fermi, I-50125, Firenze, Italy \label{arcetri}
}



  \abstract
  {We present a strong lensing analysis of the galaxy cluster PSZ1 G311.65-18.48 ($z=0.443$) using multi-band observations with Hubble Space Telescope, complemented with VLT/MUSE spectroscopic data. The MUSE observations provide redshift estimates for the lensed sources and help reducing the mis-identification of the multiple images. Spectroscopic data are also used to measure the inner velocity dispersions of 15 cluster galaxies and calibrate the scaling relations to model the subhalo cluster component. The model is based on 62 multiple images grouped in 17 families belonging to 4 different sources. The majority of them are multiple images of compact stellar knots belonging to a single star-forming galaxy at $z=2.3702$. This source is strongly lensed by the cluster to form the Sunburst Arc system. To accurately reproduce all the multiple images, we build a parametric mass model, which includes both cluster-scale and galaxy-scale components. The resulting model has a r.m.s. separation between the model-predicted and the observed positions of the multiple images of only $0.14$\arcsec. We conclude that PSZ1 G311.65-18.48 has a relatively round projected shape and a large Einstein radius (29\arcsec\ for $z_s = 2.3702$), which could indicate that the cluster is elongated along the line of sight. The Sunburst Arc source is located at the intersection of a complex network of caustics, which explains why parts of the arc are imaged with unprecedented multiplicity (up to 12 times). 
  }


   \maketitle
%

\section{Introduction}
The cores of galaxy clusters are massive and compact enough to deflect the light from distant sources in their background by several tens of arcsec via the so-called gravitational lensing effect. Due to their astigmatism, cluster lenses can split a single source into multiple images and cause enormous distortions observable in giant gravitational arcs. 
These effects characterize the strong lensing regime. The exact size of the region where we can observe them depends on several factors \citep{2004MNRAS.349..476T,2007ApJ...654..714H,2007A&A...461...25M,2010A&A...519A..90M}. In the most spectacular cases, it is of the order of a few square arcmins. This area is large enough to contain the images of several tens of distant galaxies simultaneously lensed by the clusters \citep[see, e.g.,][]{Postman_2012_clash,Lotz_2014HFF, Coe_2019,buffalo}. We can use these multiple image and arc systems to construct detailed models of the matter distribution in the cluster cores \citep{2011A&ARv..19...47K,Meneghetti_2017}.  

The usage of these models is manifold. For example, by comparing the model-derived density profiles and substructure distributions with those predicted by numerical hydrodynamical simulations, we can test the $\Lambda$CDM paradigm of structure formation \citep{2004ApJ...617L..13N,Natarajan_2007,2011A&A...530A..17M,Meneghetti_2014,Merten_2015,Jauzac_2018}. 
Discrepancies emerging from such comparison may signal missing critical ingredients in simulations or incorrect assumptions about the nature of dark matter  \citep{Newman_2013,Grillo_2015,Natarajan_2017,Meneghetti_2020}. In addition, the models enable us to use galaxy clusters as cosmic telescopes to investigate the distant universe \citep{2009ApJ...706.1201B,2012Natur.489..406Z,2014ApJ...792...76B,2013ApJ...762...32C,2017ApJ...843..129B}. Sources located near the so-called lens caustics are so highly magnified that we can resolve in their images intrinsic scales of a few tens of parsecs at redshifts 2-6  \citep{vanz_paving,johnson17,rigby17,cava18,vanz19,Vanzella_ionizing_2020}. The lens models allow us to study and characterize the properties of faint, small, and distant sources, which are the progenitors of present-day galaxies and globular clusters. In the latter systems the intense feedback of massive stars is expected to rapidly clean up the cold and dense gas in which they reside \citep[e.g.]{Calura_2015, Silich_2017}, carving elongated paths trough which thermal and ionizing energy can be radiated away. Thanks to such processes,  proto-GCs are also expected to have played a crucial role in the reionization of the Universe 
 \citep[e.g.]{Boylan_2018,2020MNRAS.493.4315M,2020MNRAS.492.4858H}.

This paper presents the strong lensing model of PSZ1 G311.65-17.48 (hereafter \CL) at $z=0.443$. This cluster was discovered through its Sunyaev-Zel'dovich effect in the Planck data \citep{planck_xxix_2014}. \cite{dahle_2016} found that the cluster hosts an exceptionally bright system of giant gravitational arcs. Subsequent studies revealed that the source of these arcs at $z=2.3702$,  dubbed ``Sunburst arc'' system, is very peculiar. \cite{RiveraT_2017} presented rest-frame ultraviolet and optical spectrometry of the arc, which shows evidence that the source is a young, star-forming galaxy that leaks Lyman-continuum (Ly-C hereafter) radiation through a perforated neutral medium. Observations with the Hubble Space Telescope (HST) show that such radiation originates from a bright, compact object that is multiply imaged twelve times in the Sunburst arc system \citep{RiveraT_2019}.  \cite{Vanzella_ionizing_2020} suggested that this object is a gravitationally bound star cluster with a stellar mass of $M_\star \lesssim 10^7\;M_\odot$. They also identified other nearby possible stellar clumps multiply imaged across the arc. 
More recently, \cite{Vanzella_bowen_2020} reported the discovery of Bowen emission arising from another source hosted in the Sunburst arc. This source is expected to have several currently un-detected counter-images. Thus, \cite{Vanzella_bowen_2020}  claimed that this source could be a transient stellar object whose other images could become detectable in the future or have dimmed out already. Alternatively, the images could be below the detection limit of the current observations. 
An accurate lens model of \CL\ is crucial to confirm the transient nature of the source detected by \cite{Vanzella_bowen_2020}  and to understand the physical properties of the many compact star-forming regions identified in the Sunburst arc system. With such a model, we could estimate the magnification of the lensed images, de-lens the arc, and even compute the time-delay surface for the transient source. 
The main goal of this paper is to illustrate the procedure to build the strong lensing model of \CL\ and discuss its properties. The characterization of the Sunburst arc source will be the focus of two other  papers (Vanzella et al., 2021; Bergamini et al., in prep.). 

The paper is organized as follows. In \Sec\ref{sec:Data} we describe the observational dataset. In \Sec\ref{sec:lens_model} we explain the method employed to construct the lens model. In \Sec\ref{sec:res} we discuss the main results obtained from our reconstruction. Finally, in \Sec\ref{sec:conclusions} we summarize our work and draw our conclusions.

Throughout this work, we adopt a flat $\Lambda$CDM cosmology
with $\Omega_m = 0.3$ and $H_0= 70\,\mathrm{km\,s^{-1}\,Mpc^{-1}}$. In this cosmological framework, a projected distance of $1\arcsec$ corresponds to a physical scale of 5.71 kpc at  $z=0.443$. All magnitudes are given in the AB system.

\section{Data}
\label{sec:Data}
In this section, we briefly describe the \CL\ photometric and spectroscopic data-sets used in our analysis. In addition, we summarize the process to construct the catalogs of multiple images and cluster members for building the cluster lens model. 
\begin{figure*}[h]
\centering
	\includegraphics[width=1\linewidth]{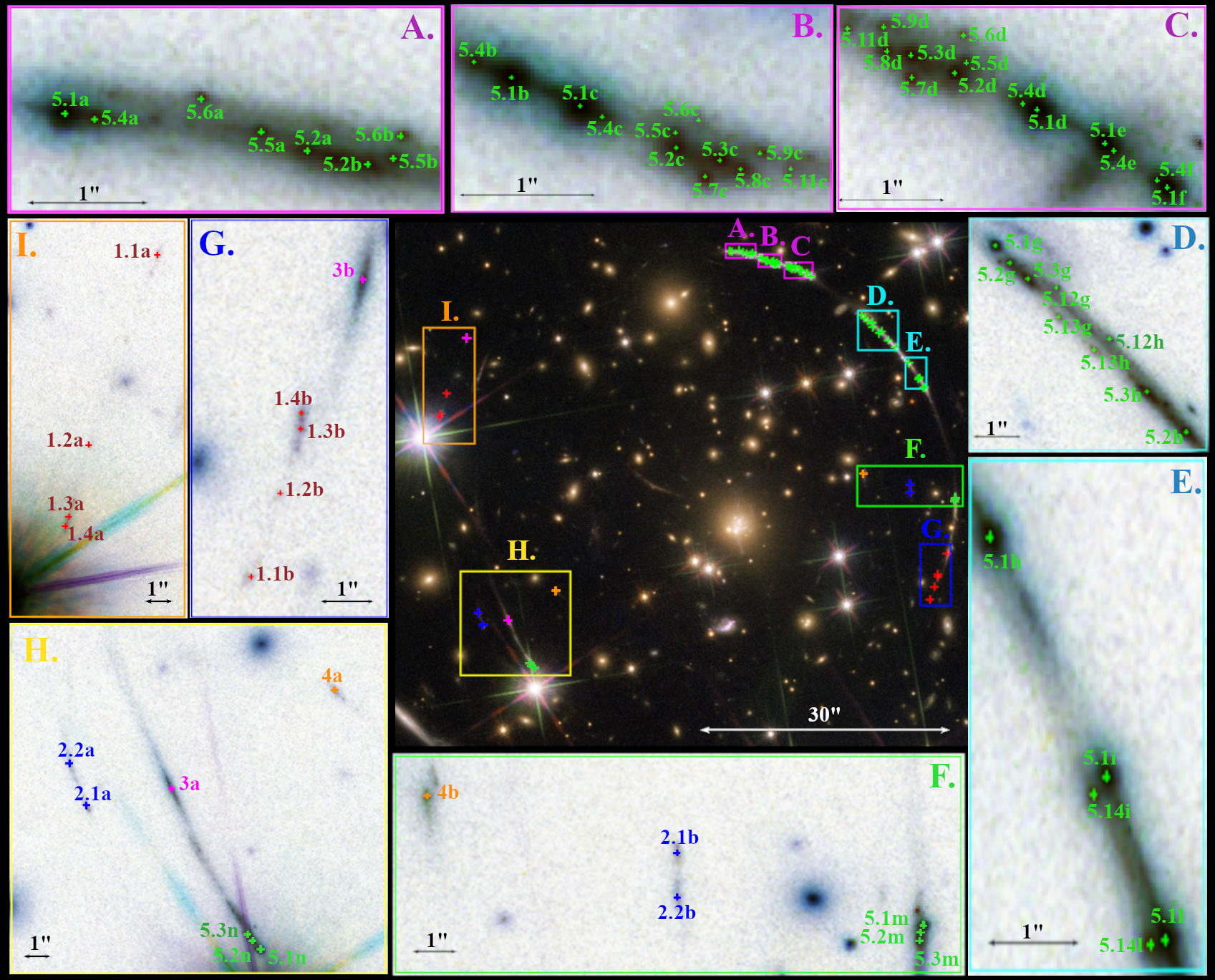}
	\caption{HST color composite image of \CL\ with all cataloged multiple images shown. The first segment of the Sunburst Arc is highlighted in the magenta cut-outs \textit{A},\textit{B} and \textit{C}, and with the cyan cut-outs \textit{D} and \textit{E}, we can see almost all families of Sys-5, marked in green. Sys-1 (in red) is visible in both the orange (\textit{I}) and blue (\textit{G}) cut-out. In the latter we can also see family 3b, marked in magenta. Sys-2 (in blue) is visible in both the yellow (\textit{H}) and green (\textit{F}) cut-out. In cut-out \textit{H} we can also see family 4a (in orange) and 3a, 5.1n, 5.2n, 5.3n. Images 4b, 5.1m, 5.2m, and 5.3m are visible in cut-out \textit{F}.}
	\label{fig:multiple}
\end{figure*}

\subsection{HST imaging and MUSE spectroscopic observations of \CL}
To perform our analysis, we use multi-band archival observations of \CL\,  carried out in the framework of several HST programs  between February 2018 and June 2019. The observations employed both the ACS and WFC3 cameras and span a wide range of wavelengths. Specifically, we use UVIS imaging in the F275W and F555W bands (program IDs 15101 and 15418 respectively, PI: H. Dahle), observations in the F814W band (program ID 15101, PI: H. Dahle), and near-IR imaging in the  F105W, F140W (program ID 15101, PI: H. Dahle) and F160W bands (program ID 15377, PI: M. Bayliss). 
More details on the procedures used for the data reduction are provided in Vanzella et al. (2021).

We complement the HST data with VLT/MUSE spectroscopic observations. The MUSE data-cube has a field-of-view of 1 $ \text{arcmin}^{2} $ with a spatial sampling of 0.2"; the wavelength range covers from 4700 \AA\ to 9350 \AA\ with a dispersion of 1.25 \AA/pix, and a spectral resolution of  $ \sim 2.6$ \AA, rather constant across the entire spectral range. These observations were acquired during May-August 2016 as part of the DDT programme 297.A-5012(A) (PI. Aghanim) with a total integration of 1.2 hours \citep{Vanzella_bowen_2020}.


\subsection{Multiple images}
\label{sec:image_cat}
The strong lensing constraints used to build the lens model of \CL\ are in the form of positions of multiple images of several sources detected in the HST images. These sources are identified using criteria based on 1) the morphology of the lensed features; 2) parity inversion rules which apply to multiple images separated by a critical line; 3) color similarities between multiple images; 4) the geometry of the lensed system; and  5) similar redshift measured from MUSE spectra. As the presence of an extended and very elongated ring-like system such as the Sunburst arc suggests, \CL\ is characterized by a projected mass distribution with a low degree of ellipticity. Galaxy clusters forming in the context of a $\Lambda$CDM model typically have prolate (i.e. cigar-like) triaxial shapes \citep{2002ApJ...574..538J,2017MNRAS.466..181D}, which may suggest that the line-of-sight to \CL\ is nearly aligned with the cluster major axis. As a result, \CL\ is a powerful gravitational lens, characterized by a large Einstein radius ($\sim 29$\arcsec\ for $z_s=2.37$) and a high magnification power \citep{2007ApJ...654..714H,2010A&A...519A..90M,Meneghetti_2014}. In some cases, the images of lensed galaxies are so highly magnified that we can resolve several compact, knot-like, star-forming regions in them. These knots are individually strongly lensed and seen multiple times in the images of their host galaxy. In our analysis, we treat each knot as an independent source belonging to the same {\em system}. The multiple images of the same knot form a {\em family}. Each family provides a constraint on the lens deflection field. Indeed, given a source at the intrinsic angular position $\vec\beta$, the positions of its images, $\vec\theta_i$, satisfy the lens equation,
\begin{equation}
    \vec\beta -\vec\theta_i = \frac{D_{LS}}{D_S}\vec\alpha(\vec\theta_i) \;,
\end{equation}
where $D_{S}$ and $D_{LS}$ are the angular diameter distances of the source and between the lens and the source respectively, and $\vec\alpha(\vec\theta)$ is the lens deflection angle at position $\vec\theta$. 

Following \cite{Bergamini_2020}, we identify the multiple images using an ID containing a number and a letter, where the integer part of the number identifies the system, the fractional part of the number identifies the family, and the letter identifies all the images belonging to the same family.

In \CL\, we find 5 candidate systems (i.e. 5 strongly lensed galaxies) at different redshifts. Their 70 multiple images are shown in \Fig\ref{fig:multiple}. In the following we briefly summarize the characteristics of these systems.
\begin{itemize}
    \item Sys-1 is indicated in red. It consists of 4 families, each with 2 multiple images. Only images [1.1-1.4]b fall within the footprint of the MUSE observations. From their spectra, we measure a system redshift $z_{1}=3.505$; 
    \item Sys-2 (shown in blue) consists of two families with two multiple images each. The redshift of this system could not be measured spectroscopically. As it will be discussed in detail later, we fit the system redshift as a free-parameter of the lens model  and we estimate it to be $ z_{2}=2.196^{+0.024}_{-0.023} $;
    \item Sys-3 and Sys-4 (indicated in magenta and orange) only contain one visible family. Each of these two systems consists of two multiple images. The sources are at redshifts $z_{3}=2.393$ and $z_{4}=1.186$, respectively;
    \item Finally, most of the strong lensing constraints are provided by the Sunburst arc system (Sys-5 at $z_{5}=2.3702$), in which we identify 54 multiple images of 13 knots, indicated with IDs 5.1-5.9, 5.11-5.14)\footnote{Image 5.10 corresponds to the candidate transient object reported in \cite{Vanzella_bowen_2020} and was not used in this paper}. Of these, family 5.1 contains twelve multiple images (5.1a-5.1n) of the LyC knot. Other knots have lower multiplicity. All the multiple images identified in the Sunburst arc system are shown in green in Fig.~\ref{fig:multiple}. 
\end{itemize}

We confirm most of the multiple image associations using the MUSE spectra. As explained earlier, in the case of Sys-1, it is not possible to measure the spectra of all images. Indeed, images [1.1-1.4]a do not fall in the footprint of the MUSE observations, as shown in Fig.~\ref{fig:CM_colors}. In addition, they are very close to a bright star, whose presence makes even more uncertain their associations to images [1.1-1.4]b based on color similarities. In fact, several tests done during the construction of the lens model of \CL\ raise doubts that our associations are correct. For example, lens models built using Sys-1 as a constraint predict additional unseen multiple images of Sys-3. For these reasons, we conclude that Sys-1 is unsecured and we prefer to exclude it form the list of constraints to build our reference model. 

As it will be discussed in details in Sect.~\ref{sec: reproduction_MI}, we use Sys-2 even without measuring its spectrum and redshift. We think that the multiple image associations are robust and well supported by the lens model. Thus, the final catalog used for the mass reconstruction of \CL\ contains 62 multiple images from 17 sources in total. The RA and DEC coordinates and the redshifts of all these images are listed in \T\ref{table:multimagesummary}. 


\subsection{Cluster members selection and measured stellar velocity dispersions}
\label{sec: data_CM}
\begin{figure}
	\centering
	\includegraphics[width=1\linewidth]{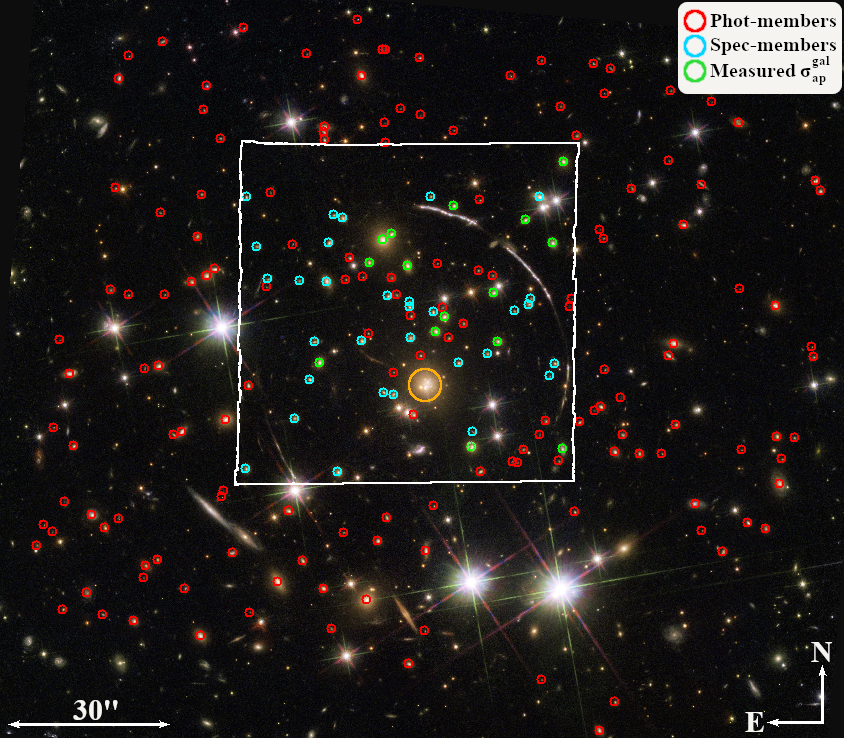}
	\caption{Color composite HST image of \CL\ with the contours of the MUSE pointing marked in white. The 197 cluster members selected in the final catalog are marked with red circles if photometrically identified (151 members), in cyan if spectroscopically confirmed (46 members), and in green (15 members) when we can measure their stellar velocity dispersion ($\sigma_{ap}^{gal}$) from the MUSE spectra. The BCG is encircled in orange.}
	\label{fig:CM_colors}
\end{figure}

Cluster galaxies are an important ingredient for the lens model, because their light is assumed to trace the cluster mass distribution. The procedure to build the catalog of these cluster members is as follows. Initially, we identify the cluster galaxies using color information derived from the HST images. Early-type cluster galaxies are expected to lay on a red-sequence in a color-magnitude diagram. We construct such diagram using the magnitudes measured in the F814W and F160W bands. We use \texttt{SExtractor} \citep{SE_BertinArnouts} in Dual Image Mode on a stack of the near-IR HST images for detecting the sources, determining their positions, and measuring their Kron-like elliptical aperture magnitudes. We select the sources near the cluster red-sequence with colors $ 0.7 < m_{F814W}-m_{F160W} < 2.0 $, and obtain an initial catalog of 273 objects brighter than  $ m_{F160W}=24$. Following a visual inspection of this first catalog, we exclude 73 sources because they are recognized to be stars. Then we proceed analyzing the MUSE datacube. We measure the redshifts for 54 sources. Of these, we consider the 46 confirmed cluster members having velocities $\pm 3000$ $\text{km s}^{-1} $ from the median cluster redshift, $z=0.4436$ (the resulting redshift range is $[0.4337, 0.4581]$). Combininig the spectroscopic and the photometric identifications, we obtain a final catalog of 197 cluster members (excluding the brightest-central-galaxy - BCG), shown in \Fig\ref{fig:CM_colors}. 

\begin{figure}
	\centering
	\includegraphics[width=1\linewidth]{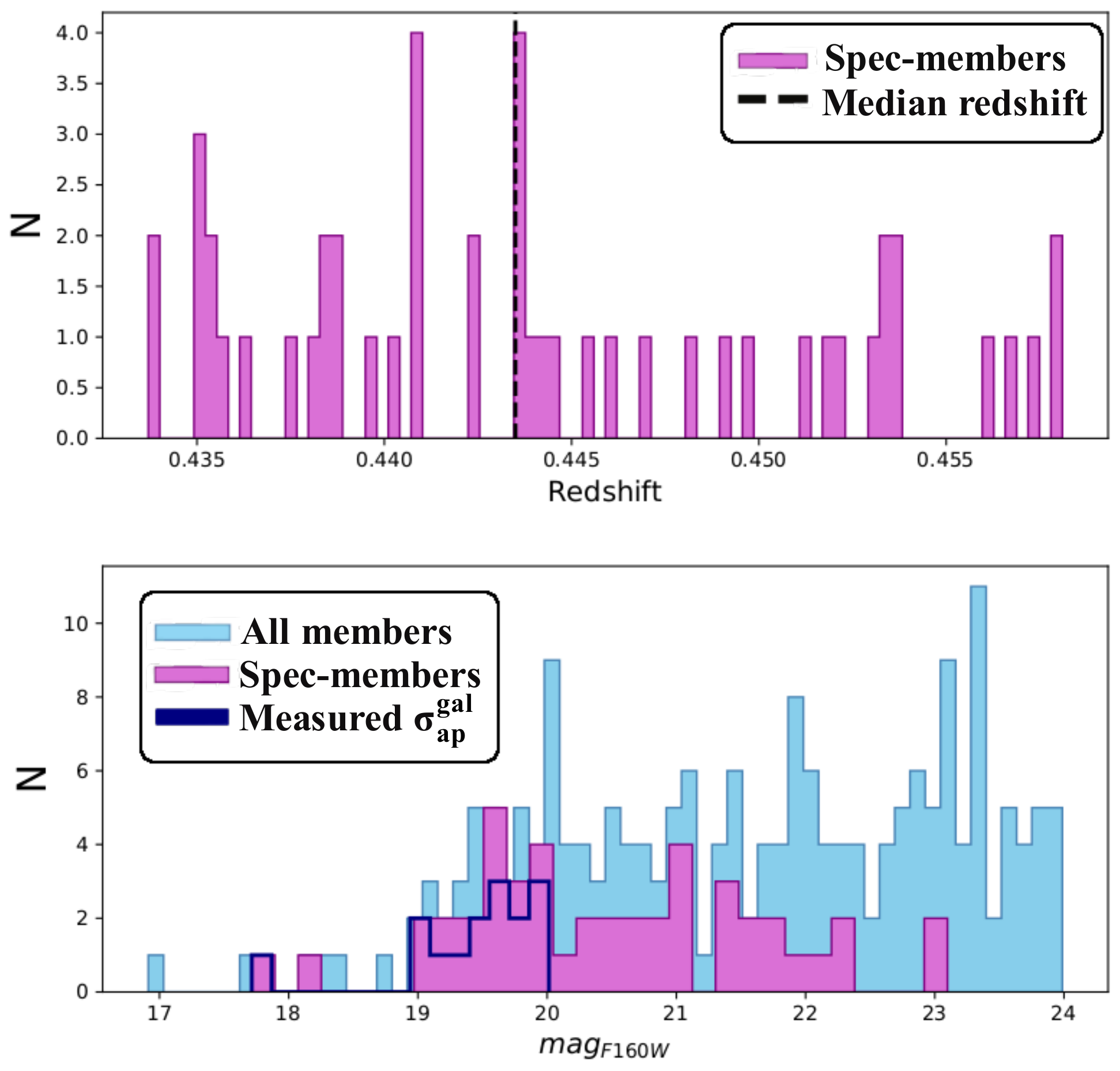}
	\caption{\textit{Upper panel:} Redshift distribution of the spectroscopically confirmed cluster galaxies with velocities within $\pm 3000$ $\text{km s}^{-1} $ from the median cluster redshift, $z=0.4436$ (black dashed line).  \textit{Bottom panel:} Magnitude distribution the cluster galaxies in the F160W band. The light-blue filled histogram shows the distribution of all 197 cluster members in our catalog, while the magenta histogram refers to the 46 spectroscopically confirmed members. The magnitude distribution of the 15 cluster members for which we could measure the velocity dispersion is given by the blue histogram.}
	\label{fig:cl_zmag}
\end{figure}

The redshift distribution of the spectroscopically confirmed cluster galaxies is shown in the upper panel of \Fig\ref{fig:cl_zmag}. The median redshift is indicated by the vertical dashed line. The F160W magnitude distribution of the same sources is given by the magenta histogram in the bottom panel of the same Figure. The light-blue filled histogram shows the magnitude distribution of all the cluster members.

\bigskip
Following the studies by \cite{Bergamini_2019, Bergamini_2020}, we measure the line-of-sight (l.o.s.) stellar velocity dispersions ($ \sigma_{ap}^{gal} $) of the 46 cluster galaxies for which we could obtain spectra. 
We extracted these spectra from the MUSE datacube using circular apertures of radius $ R_{ap}=0.8''$, which yields the best compromise between high \sn\ and low contamination from nearby bright sources. We measure the velocity dispersions by cross-correlating the observed spectra with a set of stellar templates using the \ppxf\ software \citep{Cappellari_2004, Cappellari_2017}. We perform the template fitting in the rest-frame wavelength range $[3600-5000]$\AA\, which includes all the relevant galaxy absorption lines (such as the Ca doublet), but avoids the reddest wavelengths of the MUSE spectra, which are the most contaminated by sky-line residuals. If the sky-line residuals are high, we mask the corresponding pixels to exclude them from the fitting procedure and avoid the introduction of biases in the measurements. 

\begin{figure}
	\centering
	\includegraphics[width=1\linewidth]{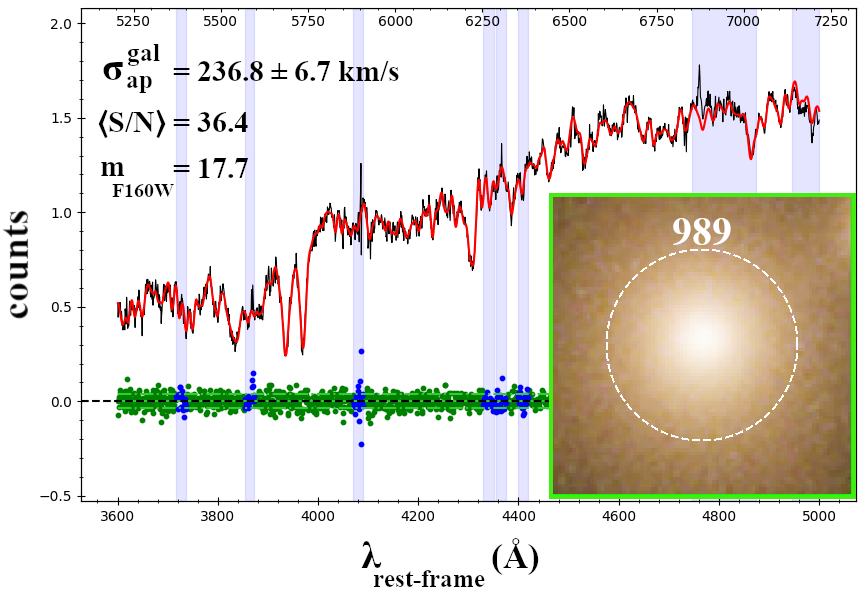}
	\caption{\ppxf\ fit of the spectrum of the cluster member \textit{Gal-}989. The RGB HST image of the galaxy is shown in the cut-out on the bottom-right corner of the Figure. The galaxy spectrum, extracted from the dashed white circular aperture shown in the cut-out, and the best-fit template are given by the black and red lines, respectively. The resulting velocity dispersion ($\sigma_{ap}^{gal}$) and \sn\ are reported in the upper-left corner, together with the F160W magnitude of the galaxy. The green and points show the fit residuals. In performing the fit, we exclude the wavelengths contained in the vertical blue bands (with corresponding residuals in blue points), due to the presence of sky-line residuals.}
	\label{fig:ppxf_fit}
\end{figure}

\begin{figure}
	\centering
	\includegraphics[width=1\linewidth]{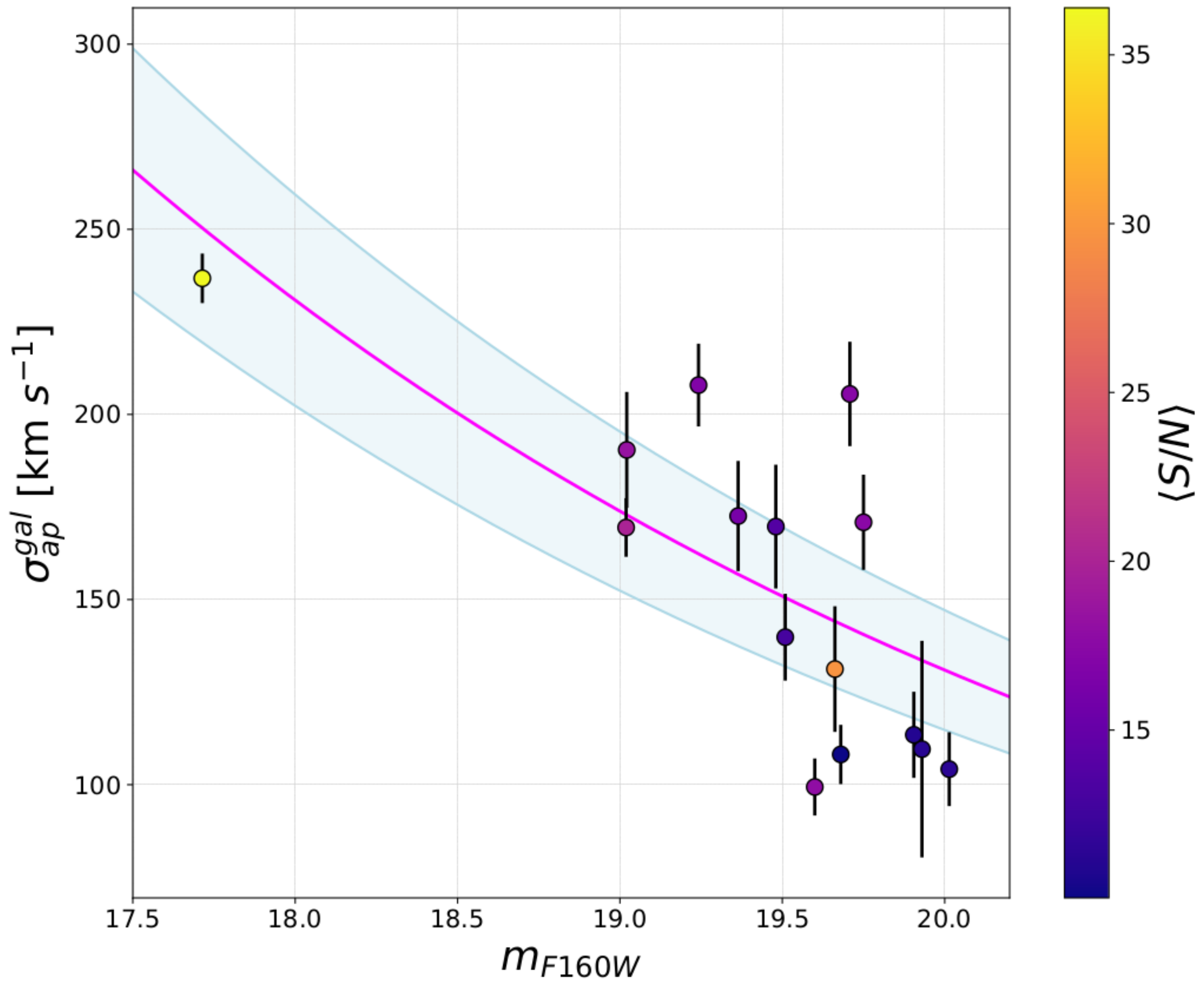}
	\caption{The measured internal stellar \sig\ of 15 cluster galaxies as a function of their F160W magnitudes (filled circles). The colors encode the mean signal-to-noise ratio of each measurement, as reported in the color-bar on the right of the Figure. The magenta line is the best-fit $\sigma-mag$ relation obtained as explained in \Sec\ref{sec: galaxy_mass_distribution}, while the light-blue colored band shows the mean scatter around the best-fit, $\Delta \sigma_{ap}$.}
	\label{fig:faberjackson}
\end{figure}

We show an example of a spectral fit performed with \ppxf\ in \Fig\ref{fig:ppxf_fit}).
For the i-th cluster member galaxy we measure the velocity dispersion,  $\sigma_{ap,i}^{gal} \pm \delta\sigma^{gal}_{ap,i}$, where $\delta\sigma^{gal}_{ap,i}$ is the associated error, and the signal-to-noise, $\left<S/N\right>_{i}$. To ensure robust measurements, we only consider galaxies with $ \sigma_{ap}^{gal}>60$ km/s and a \sn$>10$. Only 15 galaxies satisfy these selection criteria. We show how the velocity dispersions of these sources varies as a function of the F160W magnitude in \Fig\ref{fig:faberjackson}  The magnitude distribution of this sub-sample of cluster galaxies is given by the blue histogram shown in the bottom panel of \Fig\ref{fig:cl_zmag}. 

\section{Strong lensing models}
\label{sec:lens_model}

We use the public software \LT\ \citep{Kneib_lenstool,Jullo_lenstool,Jullo_Kneib_lenstool} to build the strong lensing model of \CL. Following the parametric approach implemented in this software, we decompose the cluster potential into several components, each modeled with a set of parameters. 
Let $\mathbf{p}$ be the totality of the parameters used to model all components. The cluster potential is constrained by maximizing the posterior probability distribution
\begin{equation}
    P(\vec\theta^{obs}|\mathbf{p}) \propto P(\mathbf{p}|\vec\theta^{obs})\cdot P(\mathbf{p}) \;,
    \label{eq:posterior}
\end{equation}
where $\vec\theta^{obs}$ are the observed positions of the 62 previously identified multiple images. The model likelihood is given by
\begin{equation}
    P(\mathbf{p}|\mathbf{\theta^{obs}}) \propto \exp{\left[-\frac{1}{2}\chi^2(\mathbf{p})\right]} \;.
\end{equation}
The $\chi^2(\mathbf{p})$ function quantifies the displacement between the observed multiple image and the model-predicted positions:
\begin{equation}
    \label{eq.: chi_lt}
    \chi^{2}(\mathbf{p}):=\sum_{i=1}^{N_{fam}} \sum_{j=1}^{n_{i}}\left(\frac{\left\|\vec\theta_{i, j}^{o b s}-\vec\theta_{i, j}^{p r e d}(\mathbf{p})\right\|}{\Delta \vec\theta_{i, j}}\right)^{2},
\end{equation}
where $\vec\theta_{i, j}^{o b s}$ and $\vec\theta_{i, j}^{p r e d}$ are the observed and model-predicted position of the $j$-th multiple image belonging to the $i$-th family,  $N_{fam}$ is the total number of multiple image families (in our case $N=18$, considering all four systems), and $ n_{i} $ is the number of multiple images belonging to the $i$-th family. We assume an isotropic uncertainty $\Delta{\vec\theta}_{i,j}$ on the observed positions of the images \citep{Jullo_lenstool}. We initially set this uncertainty to 0.5\arcsec.

Since the lens model is constrained by the positions of $N_{im}^{tot}=\sum_{i=1}^{N_{fam}} n_{i}$ observed multiple images, by defining $N_{par}$ as the total number of model free parameters, we can write the number of degrees-of-freedom (DoF) of the lens model as:
\begin{equation}
    \label{eq.: DoF}
    \mathrm{DoF} = 2 \times N_{im}^{tot} - 2 \times N_{fam} - N_{par} = N_{con}-N_{par} \;.
\end{equation}
The term $2 \times N_{fam}$ stems from the fact that the unknown positions of the $N_{fam}$ background sources are additional free parameters of the model. Thus, $N_{con}$ is the effective number of available constraints.

\LT\ samples the posterior distribution in Eq.~\ref{eq:posterior} using a Markov Chain Monte Carlo (MCMC) method. The uncertainties on the model parameters are determined by re-sampling the posterior distribution after re-scaling the errors on the multiple image position, $\Delta \vec\theta_{i,j}$, such that the reduced $\chi^2$ is close to 1.

We use the root-mean-square separation between the observed and model-predicted positions of multiple images, $\Delta_{rms}$ as an indicator for the goodness of our lens model \citep[see e.g.][]{Caminha_rxc2248, Bergamini_2020}:  

\begin{equation}
    \label{eq.: rms_lt}
    \Delta_{r m s}=\sqrt{\frac{1}{N_{im}^{tot}} \sum_{i=1}^{N_{im}^{tot}}\left\|\Delta_{i}\right\|^{2}}, \quad \text { with } \quad \Delta_{i}=\vec\theta_{i}^{o b s}-\vec\theta_{i}^{p r e d},
\end{equation}

\noindent where $ \Delta_{i} $ is the displacement of the $i$-th observed multiple image from the predicted position. 

\begin{table*}[h!]
	\tiny
	\def\arraystretch{1.6}
	\centering    
	\begin{tabular}{|c|c|c|c|c|c|}
	   \hline
	   \multicolumn{6}{|c|}{\textbf{Measured parameters of the scaling relations}}\\
	   \hline
	   \boldmath{$N(\sigma_{ap}^{gal})$} & \boldmath{$m_{F160W}^{ref}$} & \boldmath{$\sigma_{ap}^{ref}\ \mathrm{[km\ s^{-1}]}$} & \boldmath{$\alpha$} & \boldmath{$\Delta\sigma_{ap}\ \mathrm{[km\ s^{-1}]}$} & \boldmath{$\beta_{cut}(\gamma=0.2)$} \cr
	   \hline
	   15 & 17.71 & $249.8_{-28.8}^{+29.6}$ & $0.311_{-0.079}^{+0.076}$  & $30.8_{-6.2}^{+8.4}$ & $0.577_{-0.158}^{+0.151}$ \cr
	   \hline
	\end{tabular}
	\smallskip
    \caption{Parameters of the $\sigma_{ap}^{gal}\mbox{-}m_{F160W}$ (\Eq\ref{eq.: Scaling_relation_sigma}) and $r_{cut}^{gal}\mbox{-}m_{F160W}$ (\Eq\ref{eq.: Scaling_relation_rcut}) scaling relations obtained from the measured stellar velocity dispersions of $N(\sigma_{ap}^{gal})=15$ cluster member galaxies. For each value, we quote the median value and the $16^{th}$, $84^{th}$ percentiles of the marginalized posterior distribution of the parameter (see \Fig\ref{fig:cornerSR}). The normalization $\sigma_{ap}^{ref}$ is computed at the reference magnitude $m_{F160W}^{ref}=17.71$ of the brightest cluster member galaxy in our catalog. The slope $\beta_{cut}$ of the $r_{cut}^{gal}\mbox{-}m_{F160W}$ relation is inferred from \Eq\ref{eq.: slopes} (see text).  
    }
	\label{table:kinemacs_scaling_relations} 

\end{table*}

\subsection{Mass components}
\label{sec:mass_comp}

As explained earlier, we model \CL\ as a combination of simple parametric models, each characterized by its own gravitational potential. More specifically, we assume that the cluster total potential can be decomposed into a sum of large-scale and small-scale components \citep{Natarajan_1997}:
\begin{equation}
    \label{eq.: pot_dec}
    \phi_{tot}= \sum_{i=1}^{N_h}\phi_i^{CS}+\sum_{i=1}^{N_g}\phi_i^{SH}+ \sum_{i=1}^{N_b}\phi_i^{BK}.
\end{equation}
The $N_h$ large-scale potentials, $\phi_{i}^{CS}$, are meant to describe the cluster extended matter distribution.  The small-scale $N_g+N_b$ cluster potentials, $\phi_i^{SH}$ and $\phi_i^{BK}$, describe the matter (both dark and baryonic) in the cluster galaxies. In the following subsections, we discuss in details the approaches to model each of these potentials.



\subsubsection{Cluster-scale potentials}
\label{sec: clusterscale_mass_distribution}

The cluster-scale potentials describe both the smooth DM halo and the hot gas that constitutes the intra-cluster medium (ICM). The corresponding mass distributions extend from hundreds to thousands of kpc.   

Each cluster-scale halo is modeled with a projected dual Pseudo-Isothermal Ellipsoidal mass distribution \citep[dPIE,][]{Eliasdottir_lenstool, Limousin_lenstool, Bergamini_2019}. This model has six free-parameters, including the sky coordinates (RA and DEC), the projected ellipticity, $e=(a^2-b^2)/(a^2+b^2)$, where $a$ and $b$ are the major and minor semi-axes of the model, the position angle, $\varphi$, measured counterclockwise from the west direction, the central velocity dispersion, $\sigma_0$, and the core radius, $r_{core}$. A seventh parameter describing the dPIE density profile, the truncation radius, $r_{cut}$, is assumed to be very large (2000\arcsec) and kept fixed. 


We find that \CL\ is best described by two cluster-scale potentials. The center of the first is assumed to be within $3$\arcsec from the BCG. A second DM halo is introduced to take into account asymmetries in the mass distribution which are not well described by the central elliptical halo. This halo is left free to move in the North-East region of the cluster. We assume flat priors on all free parameters of the two halos, varying them within the limits given in the first two lines of Table~\ref{table:inout_lensing}.  

\subsubsection{Galaxy-scale potentials}
\label{sec: galaxy_mass_distribution}
\begin{figure}
	\centering
	\includegraphics[width=1\linewidth]{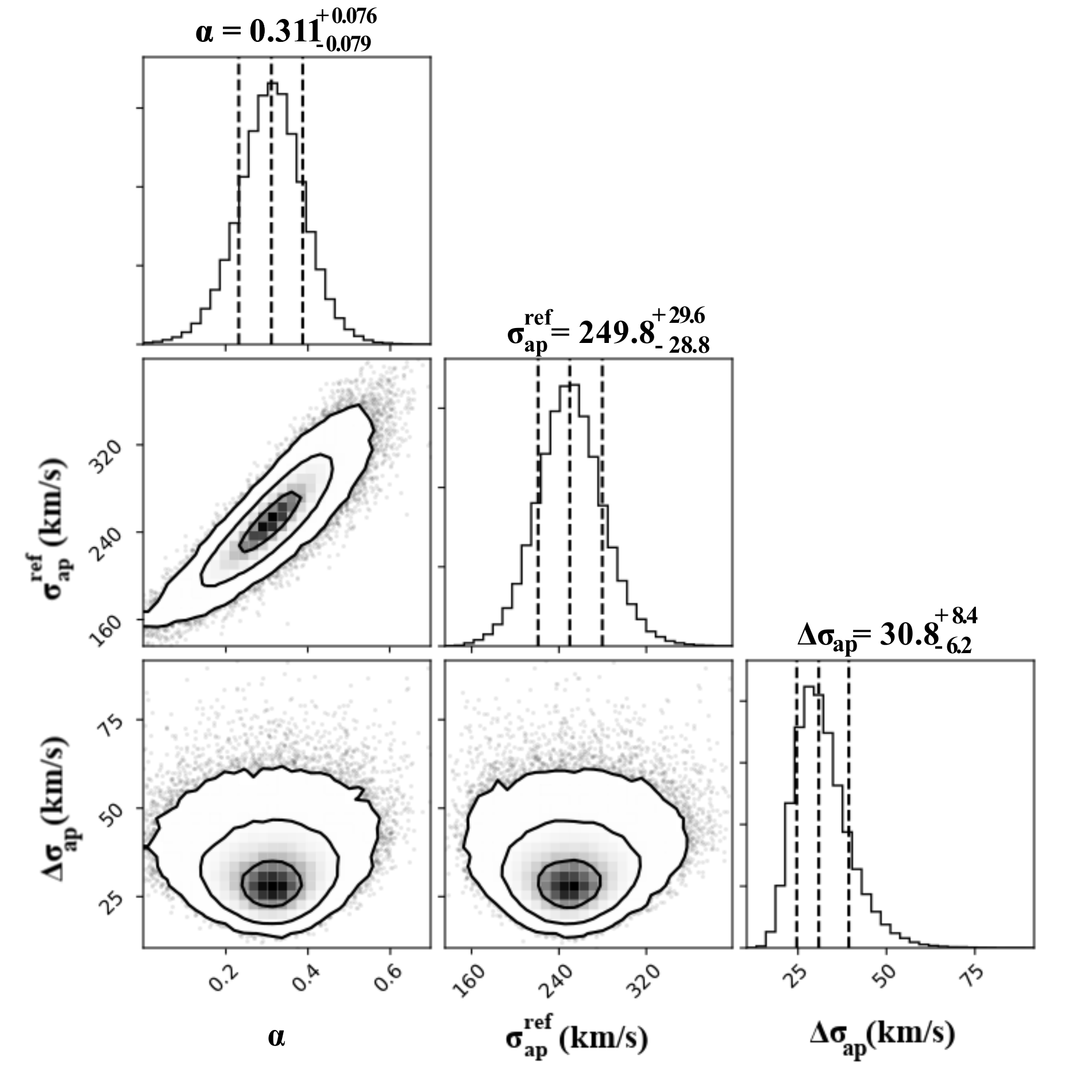}
	\caption[Degeneracy table for the scaling relation parameters]{Marginalized 2D and 1D posterior PDFs of the scaling relation parameters (slope $\alpha$, normalization $\sigma_{ap}^{ref}$, and intrinsic scatter $\Delta \sigma_{ap}$ around the Faber-Jackson relation) obtained from fitting the measured velocity dispersions of 15 cluster galaxies. The dashed vertical lines in the histograms show the median and the 16-th and 84-th percentiles of the marginalized 1D distributions of each parameter. The parameter values are reported on the top of each figure.}
	\label{fig:cornerSR}
\end{figure}
\begin{figure*}[h]
	\centering
	\includegraphics[width=1\linewidth]{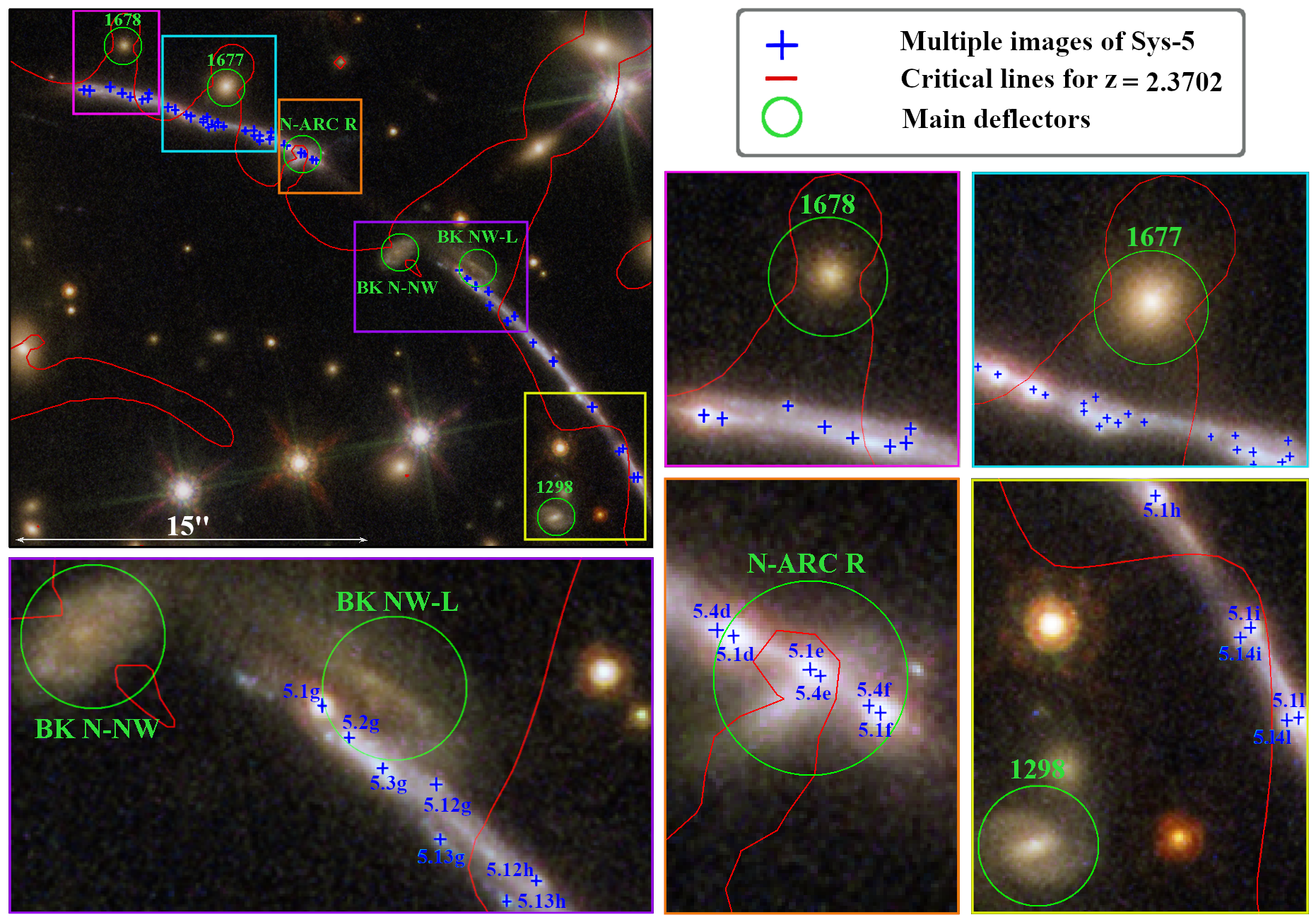}
	\caption{The upper left panel shows a composite HST image of the two northern segments of the Sunburst Arc. The positions of the main deflectors are marked with green circles of radius $0.8''$. The regions indicated with colored rectangles around them are enlarged in the other panels. The multiple images of all families of knots belonging to Sys-5 are marked in blue, while the model-predicted critical lines are shown in red.}
	\label{fig:deflectors}
\end{figure*}

In order to construct an accurate strong lensing model of \CL, the perturbing effects of cluster galaxies cannot be neglected. While the vast majority of them act to merely increase the total mass enclosed within the Einstein radius, some of them perturb the locations of multiple-images in their vicinity or alter the multiplicity and shapes of lensed images \citep{2007A&A...461...25M,Meneghetti_2020}. 


As done with the cluster-scale halo, we use dPIE models also to describe the galaxy-scale potentials. The major difficulty when modeling these sub-halos is that, while the number of cluster galaxies is high, the amount of constraints available to fit the lens model is limited. Consequently, some actions need to be taken to reduce the number of free-parameters describing these small-scale deflectors. 

For most of the cluster galaxies (i.e. those located far from the multiple images), we proceed as follows. First, we assume that the projected ellipticity of the galaxy-scale halos is zero.  \cite{Meneghetti_2017} showed that this assumption has little impact on the modeling results. We also assume that their core radii are negligible. The inner density profiles of massive early-type galaxies, such as most of the cluster members near the cluster core, are indeed consistent with singular isothermal profiles \citep[see, e.g.][]{2006ApJ...649..599K,2009ApJ...703L..51K,2010ApJ...724..511A,2018MNRAS.475.2403L}. In addition, we fix the position of each dPIE to the center of the corresponding cluster galaxy as determined by {\tt SExtractor}. Finally, following a common practice \citep[see, e.g.][]{2011A&ARv..19...47K}, we assume that the remaining parameters of the dPIE models, namely the central velocity dispersion, $\sigma_0^{gal}$, and the truncation radius, $r_{cut}^{gal}$, scale as a function of the galaxy Kron luminosity in the F160W band. Such scaling relations are implemented in \LT\ as follows:

\begin{equation}
    \sigma^{gal}_{LT}= \sigma^{ref}_{LT} \left(  \frac{L}{L_{ref}} \right)^{\alpha}\;,
    \label{eq.: Scaling_relation_sigma}
    \end{equation}
    \begin{equation}
    r^{gal}_{cut}= r^{ref}_{cut} \left(  \frac{L}{L_{ref}} \right)^{\beta_{cut}}
    \label{eq.: Scaling_relation_rcut} \;,
\end{equation}
where the first equation gives the Faber-Jackson relation \citep{Faber-Jackson_1976}.

\noindent In the above equations, $\sigma_{LT}$ is the velocity dispersion measured by \LT, which is related to the central velocity dispersion $\sigma_0$ of the dPIE model by 
\begin{equation}
    \sigma_{0}=\sqrt{\frac{3}{2}}\sigma_{LT} \;.
\end{equation}
The reference luminosity, $L_{ref}$, is that of the brightest galaxy in the cluster member catalog, which has an apparent magnitude $mag^{ref}_{F160W}=17.71$. Note that this galaxy does not coincide with the BCG. In the case of \CL, the BCG is very extended. Measuring its luminosity is difficult due to several nearby contaminating sources, including two relatively bright stars. In addition, the UV photometry shows evidence for significant star-formation activity, confirmed by the presence of strong $[\mathrm{OII}]3727$ emission lines in the MUSE spectra. Given the peculiarity of this galaxy, we prefer to exclude it from the population of the cluster members following the scaling relations in Eqs.~\ref{eq.: Scaling_relation_sigma} and \ref{eq.: Scaling_relation_rcut}. Moreover, given the lack of multiple images in the very inner region of the cluster, we find that there are not enough constraints to robustly constrain its potential. For this reason, the mass budget of the BCG is incorporated in the main cluster-scale halo.

Using the above scaling relations, the vast majority of the cluster members is described by just few parameters, namely the pivot velocity dispersion and truncation radius, $\sigma_{LT}^{ref}$ and $r_{cut}^{ref}$, and the logarithmic slopes $\alpha$ and $\beta_{cut}$. Finally, by assuming that the galaxy mass-to-light ratio scales as a function of the luminosity as $M_{tot}/L\propto L^{\gamma}$, we get rid of one additional free parameter. Indeed, the following relation holds between $\alpha$, $\beta_{cut}$, and $\gamma$:
\begin{equation}
    \beta_{cut}= \gamma -2\alpha +1.
    \label{eq.: slopes}
\end{equation}
\noindent According to the observed fundamental plane \citep{Faber_1987, Bender_1992}, we adopt $\gamma=0.2$.

As discussed in \cite{Bergamini_2019}, the cluster- and galaxy-scale mass distributions are strongly degenerate. This degeneracy can be at least partially lifted by calibrating the Faber-Jackson relation using spectroscopy. Following \cite{Bergamini_2019}, we use the velocity dispersion measurements $\sigma_{ap}$ of the 15 cluster members for which we were able to extract spectra with high \sn\ from the MUSE data-cube. As explained in Sect.~\ref{sec: data_CM}, such measurements provide the projected velocity dispersion in apertures of radius $R_{ap}=0.8$\arcsec\ centered on the cluster galaxies. We fit the $\sigma_{ap}-m_{F160W}$ relation to estimate the projected reference velocity dispersion $\sigma_{ap}^{ref}$, the intrinsic scatter $\Delta\sigma_{ap}$, and the logarithmic slope $\alpha$. We show the result of this fit with the magenta solid line in Fig.~\ref{fig:faberjackson}. The light-blue colored band shows the scatter around the best-fit. The best fit parameters are reported in Table~\ref{table:kinemacs_scaling_relations}.

In order to de-project $\sigma_{ap}$ and transform it into $\sigma_{LT}$, we use the procedure described in \cite{Bergamini_2019} (Appendix C). We then define a flat prior for $\sigma_{LT}^{ref}$ to be used with \LT. The assumed range of values is $[180\div280]$ km $s^{-1}$. The logarithmic slope $\alpha$ is fixed to the best-fit value of the $\sigma_{ap}-m_{F160W}$ relation. Finally, the parameter $\beta_{cut}$ is obtained from Eq.~\ref{eq.: slopes}.
The marginalized 1D and 2D posterior distributions for the parameters $\sigma_{ap}$, $\alpha$, and $\Delta \sigma_{ap}$, along with the median values and [16-th, 84-th] percentiles, are shown in \Fig\ref{fig:cornerSR}. 
\begin{figure*}[h]
	\centering
	\includegraphics[width=1\linewidth]{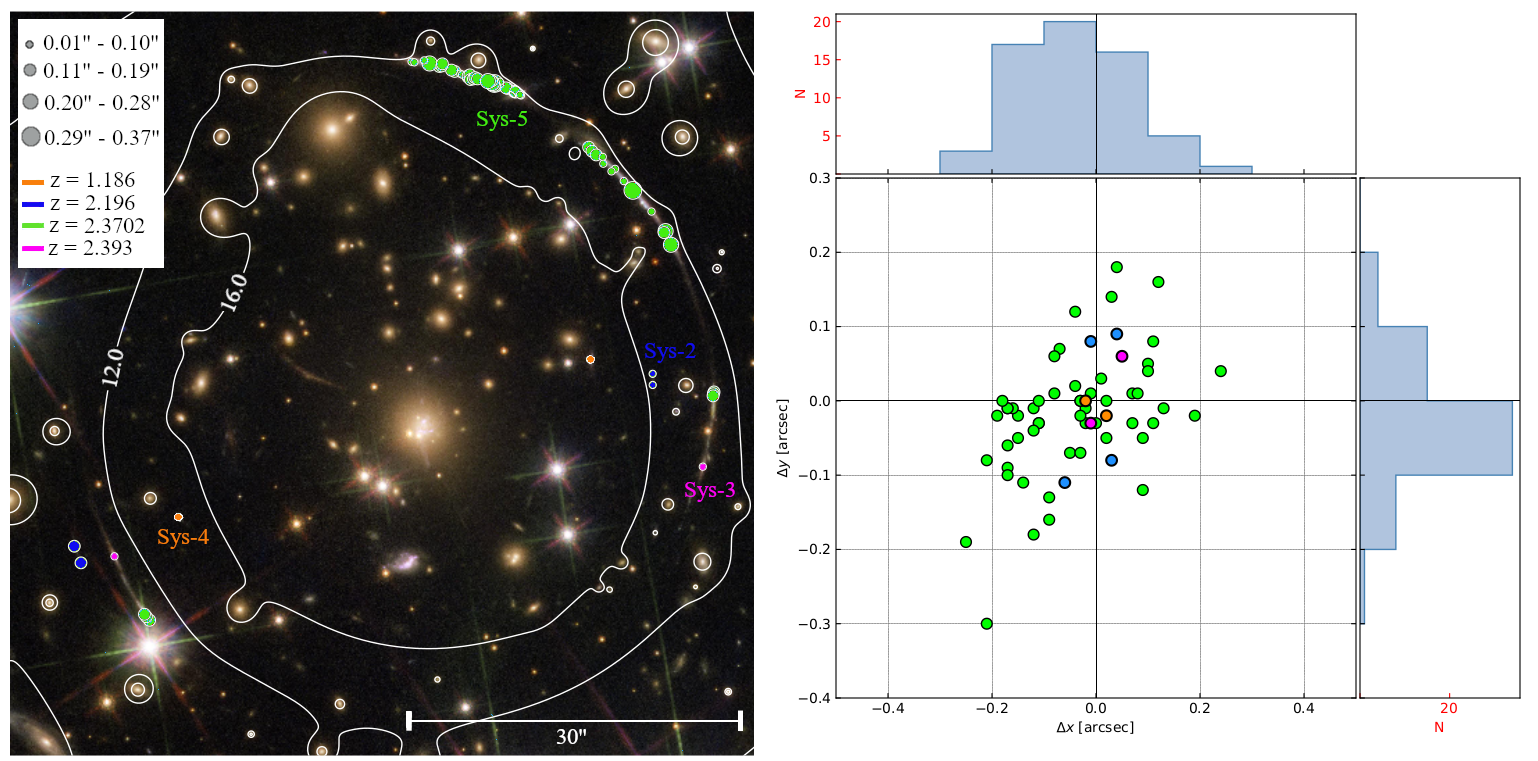}
	\caption{\textit{Left panel:} Color composite HST image of \CL\ with the observed positions of the 62 observed multiple images marked by circles. The size of the circles scales proportionally with the absolute separation ($\Vert \Delta_{i}\Vert$) between observed and model-predicted positions. The different systems are marked with different colors. Their redshifts are reported in the legend. The white overlaid contours represent the total projected mass-density distribution of our reference model in units of $10^{8}M_{\odot}/\text{kpc}^{2}$. \textit{Right panel:} The larger panel shows the distribution of the separations $\Delta x$ amd $\Delta y$ in RA and DEC between the observed and the model predicted image positions. The images are colored according to their system. The upper and right panels show the marginalized distributions of the separations $\Delta x$ amd $\Delta y$, respectively.}
	\label{fig:rms}
\end{figure*}

We use the scaling relations to model 194 of the 197 cluster members in our catalog. To model the other 3 cluster galaxies, we use a different approach, which is discussed in details in the next section. We give a summary of the parameters of the scaling relations in Table~\ref{table:kinemacs_scaling_relations}.


\subsubsection{Sunburst arc perturbers}
\label{sec: main_def}
In the case of \CL, the effects of three cluster members and three other galaxies along the line of sight to the Sunburst arc are particularly strong. 

For example, system 5.1 contains 12 multiple images, while only up to four multiple images (and a possible fifth image near the cluster BCG) would be expected from a smooth lens mass distribution. Several other knots (for example System 5.4 in sub-panel $C$ of Fig.\ref{fig:multiple}) are multiply imaged by individual  galaxies. Thus, to correctly fit the families of Sys-5 and produce a reliable mass model for \CL, we must carefully model these galaxy-scale mass components. 

In the upper left panel of \Fig\ref{fig:deflectors}, we show a view over the two northern segments of the Sunburst arc. The perturbers positions are indicated with green circles. Each perturber has a label associated. The observed multiple images of Sys-5 are marked in blue. The other panels show enlargements of the colored rectangular regions in the upper left panel.  


$Def$-1678 (RA=15:50:06.8581, Dec=-78:10:55.464) and $Def$-1677 (RA=15:50:05.4439, Dec=-78:10:57.116) are modeled as circular dPIE lenses. These deflectors have  fixed positions and core radii, while the velocity dispersions and cut radii are free to vary. The multiple images in the north-eastern and central sections of the arc, shown in the magenta and cyan rectangles, are better reproduced with these cluster members outside the scaling relations.

$Def$-1298 (RA=15:50:00.8951, Dec=-78:11:15.390) is another cluster member. Its presence is crucial to reproduce the peculiar configuration of families $5.1(h,i,l)$ and $5.14(i,l)$. These multiple images are shown in the yellow rectangle. We model this pertuber as an elliptical dPIE lens, with free-parameters given by the ellipticity, position angle, velocity dispersion and truncation radius. The position and core radius are fixed.  

$Def$-N-ARC-R (RA=15:50:04.4565, Dec=-78:11:00.191) is a possible background object. This deflector creates additional multiple images of knots $5.1(d,e,f)$ and $5.4(d,e,f)$, as shown in the orange rectangle. We model it as an elliptical dPIE, as $Def$-1298.

The remaining important pertubers are other two objects in the cluster background. $Def$-BK-N-NW (RA=15:50:03.0776, Dec=-78:11:04.142) is a galaxy at redshift $ z=0.5578 $, which we model as a circular dPIE lens optimizing its velocity dispersion and truncation radius only. The other background object, at redshift $z=0.7346$, is dubbed $Def$-BK-NW-L (RA=15:50:02.0286, Dec=-78:11:04.748). \cite{2020MNRAS.491.4442L} recently studied the circumgalactic medium of this galaxy using the gravitational arc-tomography technique \citep{lopez_2018}. We use an elliptical dPIE lens model to describe the effects of this perturber on several nearby knots in the Sunburst arc. The position of $Def$-BK-NW-L is left free to vary within $4''$ to reproduce the multiple image positions of knots $5.12(g,h)$ and $5.13(g,h)$, as shown in the purple rectangle. 

For all the model parameters, we report the initial values and the priors in the upper part of Table~\ref{table:inout_lensing}.

\section{Results}
\label{sec:res}

In this section, we show the results of the model optimization. The model presented here has grown through several intermediate steps, where an increasing number of constraints has been used and the model complexity has been gradually increased. More details about how we converged to the final lens model are given in the Appendix.


\subsection{Reproduction of the multiple images}\label{sec: reproduction_MI}

We quantify the robustness of the lens model of \CL\ using the separation between the observed multiple image and model-predicted positions.



For each system used, we show such separations along the $x-$ and $y-$ axes in the right panel of  \Fig\ref{fig:rms}. The scatter plot and the histograms on the top and on the right show that most multiple images have separations smaller than $0.2$\arcsec in both directions. Indeed, the resulting $\Delta_{rms}$ is $0.14$\arcsec. 

In the scatter plot, we use different colors to identify the multiple images by the family they belong to. Among them, families 4, 5.12, and 5.13 are the best reproduced by the model with individual r.m.s. separations of $0.02-0.03$\arcsec. Family 5.1, which has the largest number of multiple images, has a r.m.s. separation of $0.13$\arcsec. We measure the largest discrepancy between model-predicted and observed image positions for families 5.8, 5.11, and 5.14, for which the r.m.s. separation is $0.19-0.2$\arcsec.

In the left panel of Fig.~\ref{fig:rms}, we show the RGB color composite image of \CL\ with overlaid the observed multiple image positions indicated with circles. These circles are colored according to the systems they belong to. Their radii reflect the magnitude of the absolute separation between observed and model-predicted image positions, given by $\left\|\Delta_{i}\right\|$. Although most of the constraints are located in the North-West region of the cluster, from the image separations we do not find evidence that the quality of the model degrades in specific areas of the lens. As expected, the multiple images closest to the galaxy-scale perturbers have the highest measured separations, but, even in those cases, they are $\lesssim 0.35$\arcsec. The multiple images of Sys-2, 3, and 4 are very well reproduced by the model. These results are in the range of similar high-precision lens studies \citep[see e.g.][]{Caminha_macs0416, Caminha_2019}. 

Note that the model is unconstrained near the center and in the North-East region of the cluster. As explained in Sect.~\ref{sec: data_CM}, the candidate system Sys-1, whose images [1.1-1.4]a are located in this last region, cannot be accommodated into the model. Being the associations of its multiple images highly uncertain, we prefer to exclude them from the fit. More extended and deeper MUSE observations of this system will help to better constrain the cluster mass distribution in this region.   

\subsection{The case of Sys-2}\label{sec:caseofsys2}
Sys-2 deserves a dedicated discussion. Due to its faintness, we are not able to measure the redshift of its multiple images, neither to confirm their associations using spectroscopy. However, the system is easily reproduced by the lens model. Based on the location of the images with respect to those of the other lensed sources, we assume that the system redshift is in the range $1.9\div2.5$. Using this prior and fitting the image positions and redshifts with \LT, we estimate that the source originating this system is at redshift $ z_{2}=2.196^{+0.024}_{-0.023} $. In particular, we obtain very similar redshift estimates for the two families identified as belonging to the system. Indeed, for family 2.1 and 2.2 we obtain $z_{2,1}=2.197^{+0.026}_{-0.025}$ and $z_{2,2}=2.194^{+0.022}_{-0.021}$, respectively.

\begin{figure}[h]
    \centering
    \includegraphics[width=1\linewidth]{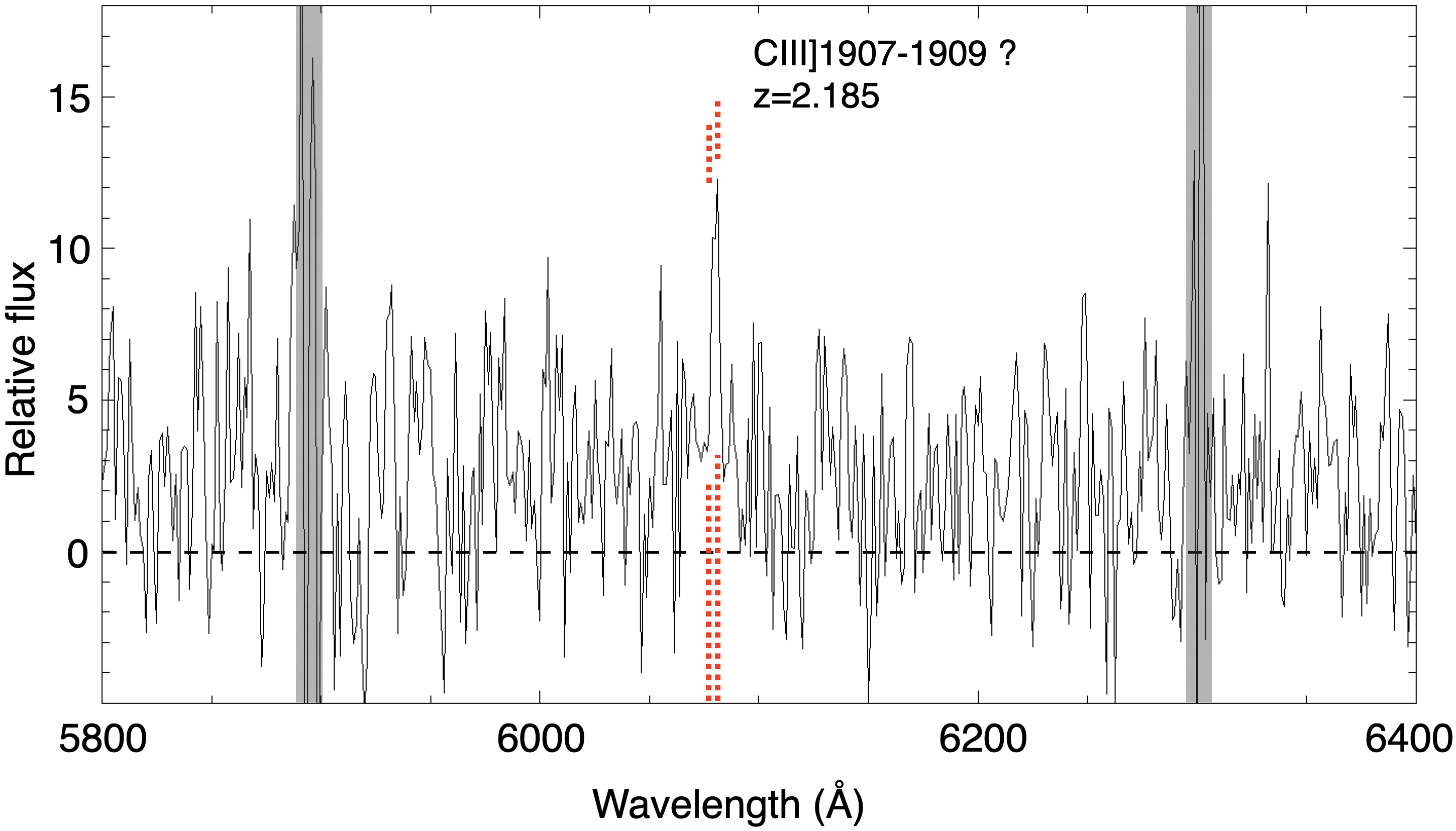}
    \caption{Stack of the spectra (in black) extracted from the MUSE datacube at the multiple image positions of Sys-2. We find hints for the presence of an emission line at $\sim 6080\AA$, which could be interpreted as a CII]1909 doublet (red dotted lines) at $z=2.185$, consistent with the redshift estimate obtained with \LT. The grey vertical bands mark the positions of known sky lines.} 
    \label{fig:sys2spec}
\end{figure}

Once estimated these redshifts, we repeated the analysis of the MUSE spectra extracted at the observed positions of the multiple images. We stacked them in order to increase the signal-to-noise ratio, obtaining the spectrum shown in Fig.~\ref{fig:sys2spec}. In this spectrum, we find hints for the presence of a possible emission line at $\sim 6080\AA$, which we interpret as a CIII]1909 doublet a $z=2.185$. This line is  consistent with the redshift estimate of Sys-2 obtained with \LT.  Based on this evidence and on the low r.m.s. separation of the multiple images ($\lesssim 0.1$\arcsec), we decide to use Sys-2 to derive the lens model of \CL.

\begin{figure}[h]
    \centering
    \includegraphics[width=1\linewidth]{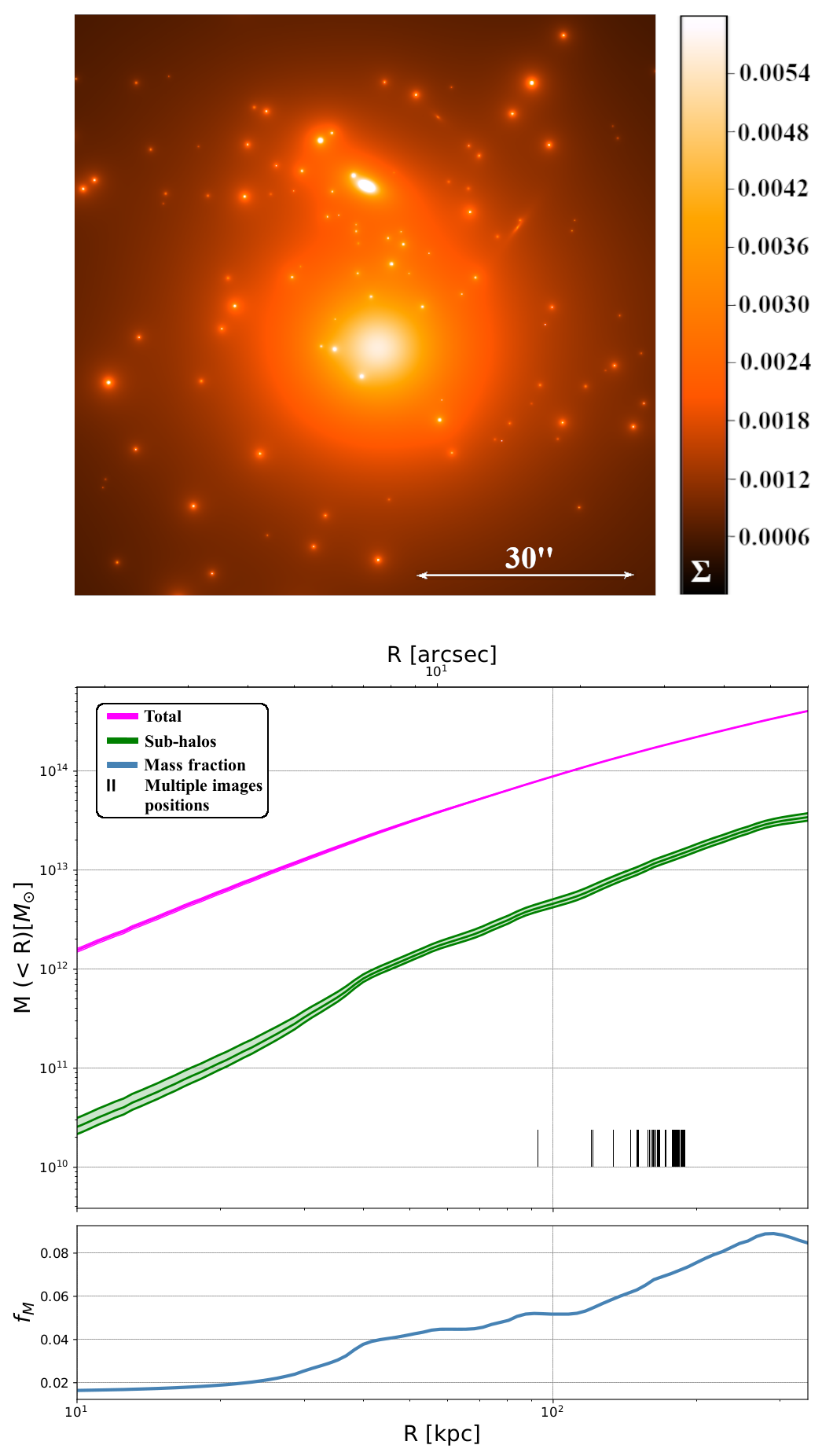}
    \caption{\textit{Upper panel}: Projected mass-density map for the reference model. The color scale is set between 0 and 0.006 in units of $10^{12}M_{\odot}/\text{kpc}^{2}$.
    \textit{Bottom panel}: Projected cumulative mass profiles of \CL\ as a function of the projected distance from the BCG ($R$). In magenta we show the median value and the 1-$\sigma$ confidence levels for the total mass profile. In green we show the median value and the 1-$\sigma$ confidence levels for the subhalo component mass profile. The positions of the multiple images are marked with vertical black segments. The fractional contribution of the cluster member galaxies mass to the total cumulative mass is shown in blue.} 
    \label{fig:massprofile}
\end{figure}
\begin{table*}
	\tiny
	\def\arraystretch{2.3}
	\centering          
	\begin{tabular}{c|c|c|c|c|c|c|c|}

\cline{2-8}
\multicolumn{1}{l|}{} & \multicolumn{7}{c|}{\textbf{Input parameter values and intervals of reference lens model}}\\
\cline{2-8}
\multicolumn{1}{l|}{} & \boldmath{$x\, \mathrm{[arcsec]}$} & \boldmath{$y\, \mathrm{[arcsec]}$} & \boldmath{$e$} & \boldmath{$\theta\ [^{\circ}]$} & \boldmath{$\sigma_{LT}\, \mathrm{[km\ s^{-1}]}$} & \boldmath{$r_{core}\, \mathrm{[arcsec]}$} & \boldmath{$r_{cut}\, \mathrm{[arcsec]}$} \\ 
          \hline
\multicolumn{1}{|c|}{\textbf{Main DM Halo}}      & -3.0 $\div$ 3.0        & -3.0 $\div$ 3.0         & 0.0 $\div$ 0.3 & -40.0 $\div$ 40.0                       & 800.0 $\div$ 1100.0                                & 2.0 $\div$ 15.0                  & 2000.0                         \\
\multicolumn{1}{|c|}{\textbf{Nord DM Halo}}      & -16.0 $\div$ 0.0       & 10.0 $\div$ 34.0        & 0.0 $\div$ 0.9 & -90.0 $\div$ 90.0                       & 300.0 $\div$ 800.0                                 & 0.01 $\div$ 15                   & 2000.0                         \\ \hline
\multicolumn{1}{|c|}{\textbf{1678}}              & 0.72                   & 34.72                   & 0.0            & 0.0                                     & 30.0 $\div$ 120.0                                  & 0.01                             & 0.1 $\div$ 15.0                \\
\multicolumn{1}{|c|}{\textbf{1677}}              & 5.04                   & 33.0                    & 0.0            & 0.0                                     & 50.0 $\div$ 150                                    & 0.01                             & 0.1 $\div$ 20.0                \\
\multicolumn{1}{|c|}{\textbf{N-ARC R}}       & 8.07                   & 29.91                   & 0.0 $\div$ 0.9 & 0.0 $\div$ 140.0                        & 10.0 $\div$ 150.0                                  & 0.01                             & 0.0 $\div$ 20.0                \\
\multicolumn{1}{|c|}{\textbf{BK N-NW}}           & 12.31                  & 25.96                   & 0.0            & 0.0                                     & 30.0 $\div$ 150.0                                  & 0.01                             & 0.1 $\div$ 15.0                \\
\multicolumn{1}{|c|}{\textbf{BK NW L}}        & 12.0 $\div$ 16.0       & 22.0 $\div$ 26.0        & 0.0 $\div$ 0.9 & -90.0 $\div$ 90.0                       & 10.0 $\div$ 200.0                                  & 0.01                             & 0.1 $\div$ 15.0                \\
\multicolumn{1}{|c|}{\textbf{1298}}              & 19.0                   & 14.71                   & 0.6 $\div$ 0.9 & 0.0 $\div$ 180.0                        & 10.0 $\div$ 200.0                                  & 0.01                             & 0.0 $\div$ 15.0                \\ \hline
\multicolumn{1}{c}{}
          \\[-5ex]
\hline
\multicolumn{1}{|c|}{\textbf{Scaling relations}} & $N_{gal}=194 $         & $m_{F160W}^{ref}$=17.71 & $\alpha$=0.31  & $\sigma_{LT}^{ref}=$ 180.0 $\div$ 280.0 & $\beta_{cut}$=0.57                                  & $r_{cut}^{ref}=$ 0.1 $\div$ 25.0 & $\gamma$=0.2                   \\ \hline
\end{tabular}
	\\[6ex]
\begin{tabular}{c|c|c|c|c|c|c|c|}
\cline{2-8}
\multicolumn{1}{l|}{} & \multicolumn{7}{|c|}{\textbf{Optimized output parameters of reference lens model}}\\
\cline{2-8}
\multicolumn{1}{l|}{} & \boldmath{$x\, \mathrm{[arcsec]}$} & \boldmath{$y\, \mathrm{[arcsec]}$} & \boldmath{$e$} & \boldmath{$\theta\ [^{\circ}]$} & \boldmath{$\sigma_{LT}\, \mathrm{[km\ s^{-1}]}$} & \boldmath{$r_{core}\, \mathrm{[arcsec]}$} & \boldmath{$r_{cut}\, \mathrm{[arcsec]}$} \\ 
          \hline
\multicolumn{1}{|c|}{\textbf{Main DM Halo}}      & $0.12_{-0.20}^{+0.17}$  & $-2.61_{-0.28}^{+0.39}$ & $0.10_{-0.01}^{+0.01}$ & $-1.11_{-2.98}^{+3.10}$                        & $942.73_{-15.13}^{+21.90}$                         & $5.16_{-0.34}^{+0.33}$                & 2000.0                         \\
\multicolumn{1}{|c|}{\textbf{Nord DM Halo}}      & $-3.27_{-0.77}^{+0.81}$ & $17.80_{-1.35}^{+1.57}$ & $0.41_{-0.07}^{+0.08}$                                          & $-26.38_{-5.03}^{+3.38}$                    & $448.36_{-40.48}^{+31.77}$                         & $1.50_{-0.83}^{+1.02}$                & 2000.0                         \\ \hline
\multicolumn{1}{|c|}{\textbf{1678}}              & 0.72                    & 34.72                   & 0.0                                                             & 0.0                                           & $90.22_{-6.06}^{+5.77}$                          & 0.01                                  & $12.99_{-2.22}^{+1.37}$         \\
\multicolumn{1}{|c|}{\textbf{1677}}              & 5.04                    & 33.0                    & 0.0                                                             & 0.0                                           & $116.57_{-6.05}^{+7.31}$                          & 0.01                                  & $8.77_{-5.24}^{+6.54}$         \\
\multicolumn{1}{|c|}{\textbf{N-ARC R}}       & 8.07                    & 29.91                   & $0.33_{-0.17}^{+0.18}$                                            & $88.31_{-44.72}^{+36.18}$                     & $61.98_{-3.22}^{+4.04}$                           & 0.01                                  & $4.56_{-1.85}^{+2.05}$         \\
\multicolumn{1}{|c|}{\textbf{BK N-NW}}           & 12.31                   & 25.96                   & 0.0                                                             & 0.0                                           & $59.55_{-14.67}^{+16.49}$                          & 0.01                                  & $10.79_{-5.17}^{+2.90}$         \\
\multicolumn{1}{|c|}{\textbf{BK NW L}}        & $12.95_{-0.52}^{+0.51}$ & $24.47_{-0.27}^{+0.30}$ & $0.64_{-0.12}^{+.12}$                                           & $23.02_{-18.83}^{+20.62}$                     & $96.10_{-16.83}^{+17.69}$                         & 0.01                                  & $5.81_{-2.04}^{+1.67}$         \\
\multicolumn{1}{|c|}{\textbf{1298}}              & 19.0                    & 14.71                   & $0.89_{-0.01}^{+0.01}$                                          & $54.63_{-1.11}^{+1.10}$                      & $93.23_{-3.67}^{+6.29}$                         & 0.01                                  & $10.20_{-3.60}^{+2.92}$        \\ \hline
\multicolumn{1}{c}{}
          \\[-5ex]
\hline
\multicolumn{1}{|c|}{\textbf{Scaling relations}} & $N_{gal}$=194           & $m_{F160W}^{ref}$=17.71 & $\alpha$=0.31                                                   & $\sigma_{LT}^{ref}= 200.56_{-8.00}^{+10.21}$ & $\beta_{cut}$=0.57                                  & $r_{cut}^{ref}=15.15_{-2.39}^{+3.21}$ & $\gamma$=0.2                   \\ \hline
\end{tabular}
	
	\smallskip
    \caption{
    {\it Top:}Input parameters values of the reference model. The single numbers are fixed parameters, while for the free parameters we report the lower and upper limits of the input flat priors. In the last row, we report the input parameters for the scaling relations used to parametrize the subhalo component.
    {\it Bottom:} Output parameters values from the optimization of the reference model. We report the median value of the parameter with errors corresponding to the [16-th, 84-th] percentiles of the marginalized posterior distribution.
    }    

	\label{table:inout_lensing}

\end{table*}
 
\subsection{Mass distribution of \CL}

The summary of the lens model parameters is given in the bottom part of Table~\ref{table:inout_lensing}. The surface density map of \CL\ is shown in the upper panel of  Fig.\ref{fig:massprofile}. We also show some surface density contours corresponding to the levels $8$, $12$, and $16\times 10^{8}\;M_\odot$ kpc$^{-2}$ overlaid to the RGB image of \CL in Fig.~\ref{fig:rms}.

As anticipated earlier, the cluster mass distribution is dominated by a rather round ($e=0.1$) main cluster-scale mass component, characterized by a velocity dispersion $\sigma_{LT}=942.73^{+21.90}_{-15.13}$ km $s^{-1}$. In order to correctly reproduce the Sunburst arc upper segment, an additional mass component is required North-East of the BCG. Its velocity dispersion is $\sigma_{LT}=448.36^{+31.77}_{-40.48}$ km $s^{-1}$. This second halo is elongated almost tangentially to the Sunburst arc and its center is not associated with any obvious cluster galaxy. We believe that this mass component accounts for some degree of asymmetry in the large-scale mass distribution of \CL. As shown by \cite{2007A&A...461...25M}, such asymmetries can have a significant impact on the cluster strong lensing properties. 

The population of the cluster galaxies is described by scaling relations whose logarithmic slopes are $\alpha=0.31$ and $\beta_{cut}=0.57$. These values agree well with those previously reported for clusters at similar redshift by \cite{Bergamini_2019} and \cite{Bergamini_2020}, who measured $\alpha \sim 0.3$ and $\beta_{cut}\sim 0.6$.
Using these logarithmic slopes, the normalization of the Faber-Jackson relation derived with \LT\ is $\sigma_{LT}^{ref}=200.56^{+10.21}_{-8.00}$. 
 
The \LT\ velocity dispersion estimates for the perturbers of the Sunburst arc are in the range $[59\div 117]$ km $s^{-1}$. The most massive perturber is $Def-1677$, for which we measure $\sigma_{LT}=116.57^{+7.31}_{-6.05}$. Converting $\sigma_{LT}$ into $\sigma_0$ and accounting for projection effects, we obtain a projected velocity dispersion of $\sigma_{ap}^{LT}\sim 130$ km $s^{-1}$. For this cluster galaxy, we were able to obtain also a spectroscopic measurement of the velocity dispersion, $\sigma_{ap}=109.6\pm29.2$ km $s^{-1}$. Given the quoted $1-\sigma$ uncertainties, the \LT\ and the spectroscopic velocity dispersion measurements agree well. This result represents another important validation test for the model on small scales.

In the bottom panel of Fig.~\ref{fig:massprofile}, we show in magenta the total mass profile in the inner region of \CL. The center is assumed to coincide with the location of the BCG. The distances of the multiple images from the cluster center are indicated by the vertical black sticks on the bottom of the upper sub-panel. Within the strong lensing region of the cluster, which extends to $\sim 200$ kpc from the center, we measure a total mass of $\sim 2\times 10^{14}\;M_\odot$. The green line shows the profile of the mass in sub-halos. The mass fraction in subhalos as a function of the distance from the center is shown by the blue curve in the bottom sub-panel. We find that the fraction of mass in subhalos is always below $10\%$ in the inner region of the cluster. Note that the BCG is not included in the subhalo mass budget. For both the total and the sub-halo mass profiles, we show the 1-$\sigma$ uncertainties using color bands, computed by extracting parameters from the MCMC chains of 500 random realizations of the model.

\begin{figure*}
    \centering
    \includegraphics[width=1\linewidth]{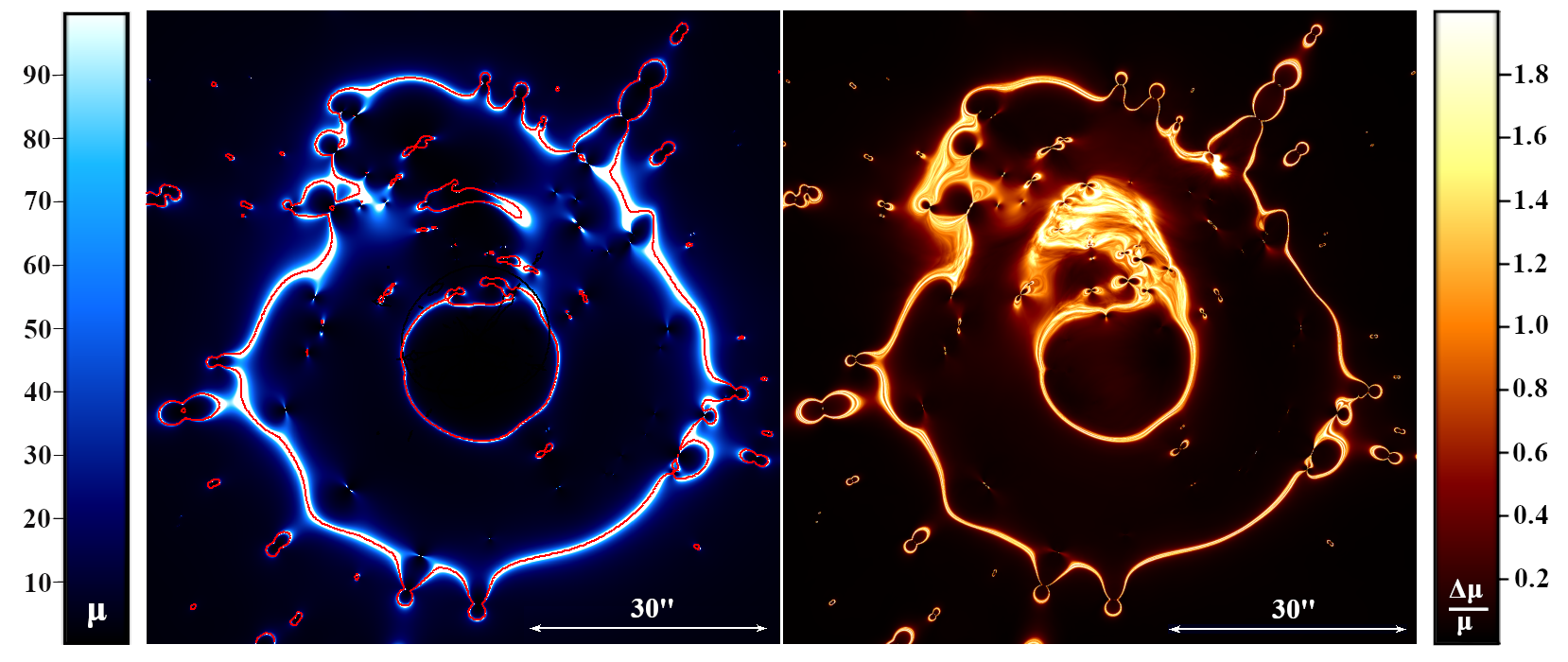}
    \caption{\textit{Left panel:} Absolute magnification map of \CL. We assume a source redshift of $z_s=2.3702$, coinciding with the redshift of the Sunburst arc. The magnification values start at $\mu=0$ and saturate at $\mu=100$. The critical lines for the same source redshift are shown in red. \textit{Right panel}: Map of the magnification uncertainties. We show the relative statistical error on the absolute magnification ($\Delta\mu/\mu$).}
    \label{fig:magnification_refmodel}
\end{figure*}

\subsection{Magnification map, critical lines, and caustics}

Finally, in the left panel of Fig.~\ref{fig:magnification_refmodel} we show the absolute magnification map of \CL. We assume a source redshift of $z_s=2.3702$, coinciding with that of the Sunburst arc. The map saturates at $\mu=100$, indicated by the white regions. 

\begin{figure}
    \centering
    \includegraphics[width=1\linewidth]{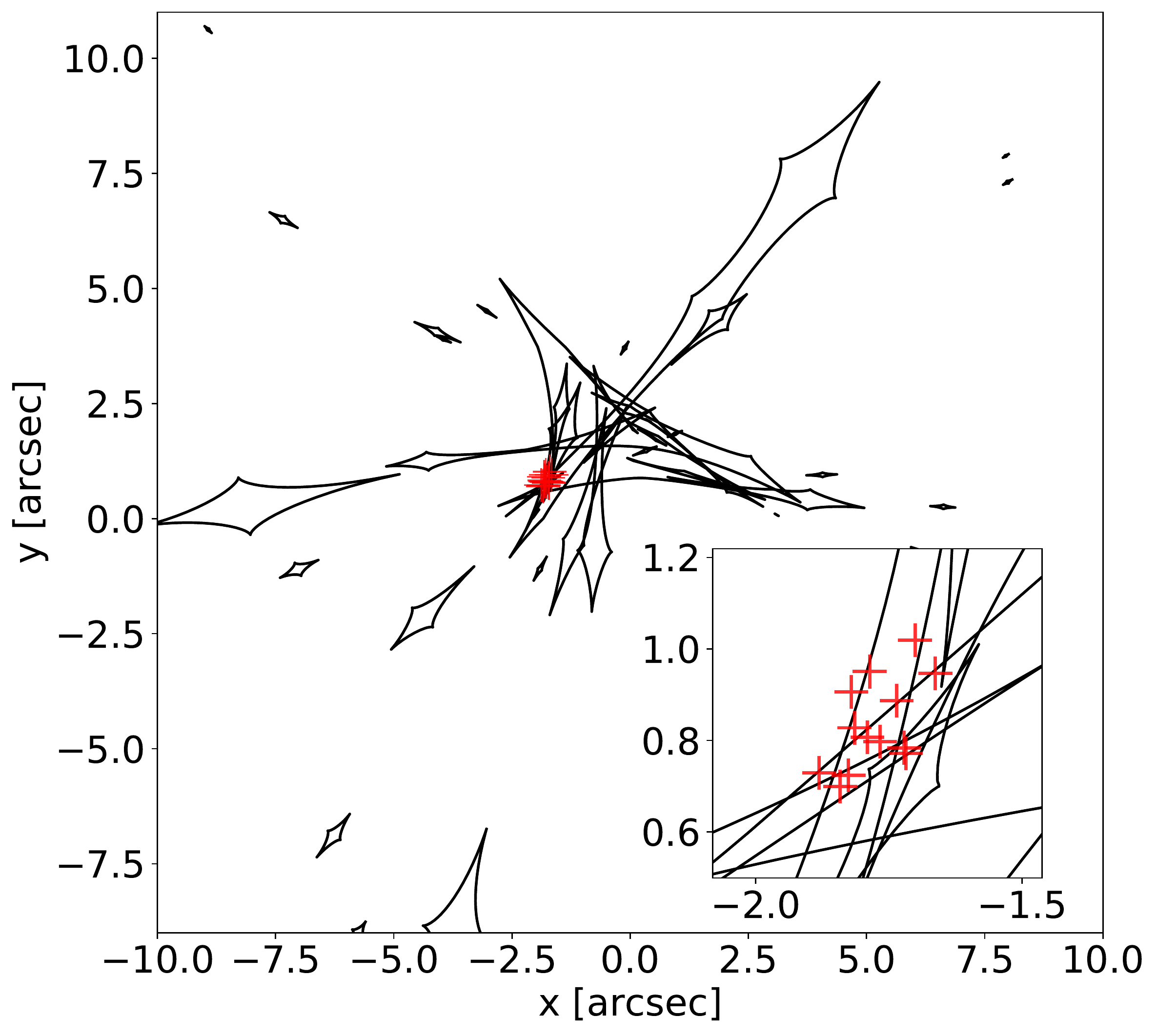}
    \caption{Caustics of \CL\ for a source redshift $z_s=2.3702$. The positions of all multiple images of Sys-5 mapped onto the source plane are shown with red crosses. The inset on the bottom-right corner shows a zoom-in on the region containing all the crosses.}
    \label{fig:caustics}
\end{figure}

The magnification is infinite along the lens critical lines, which are indicated in red. The same critical lines are also shown in Fig.~\ref{fig:deflectors},  overlaid to the HST color-composite image of the cluster. The cluster has a main tangential critical line along which the Sunburst arc is stretched and strongly magnified. This arc is likely originated from a source nearly perfectly aligned with the cluster center. As discussed earlier, it is very likely that the cluster has a prolate triaxial shape and is oriented with its major axis pointing toward us. Due to its nearly circular projected shape, the smooth cluster-scale mass component produces a roundish tangential critical line, which corresponds to a small tangential caustic on the source plane. This caustic is susceptible to perturbations by additional mass components located near the tangential critical line. These perturbers generate a complex network of resonant caustics, which is shown in Fig.~\ref{fig:caustics}. In the regions where these caustics overlap, the image multiplicity can be very high, thus explaining the high number of multiple images belonging to several families of Sys-5. The positions of the multiple images of the knots in the Sunburst arc mapped onto the source plane are given by the red crosses in Fig.~\ref{fig:caustics}. The resonant caustics correspond to the wiggles of the tangential critical line around the individual perturbers.

The right panel of Fig.~\ref{fig:magnification_refmodel} shows a map of the magnification uncertainties corresponding to the absolute magnification map shown in the left panel. We show the relative error on the absolute magnification ($\Delta\mu/\mu$). The magnification uncertainties ($\Delta\mu$) are computed as follows. We randomly extracted 100 parameter samples from the MCMC chains of the model realizations. $\Delta\mu$ is computed as the half the difference between the 84-th and 16-th percentiles, and $\mu$ is defined as the median value of the magnification distributions. The errors are particularly large close to the critical lines \citep[see ][]{Meneghetti_2017} and in the north-east region of the lens plane, where very few constraints are available to build the model.

\begin{figure}
    \centering
    \includegraphics[width=1\linewidth]{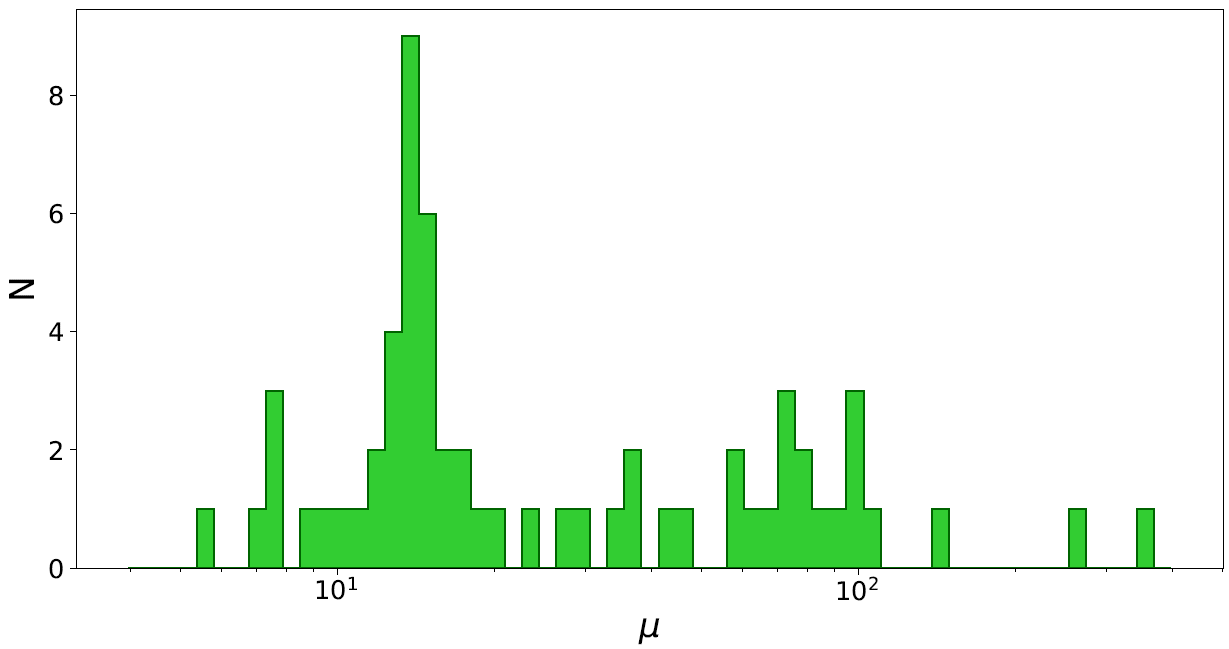}
    \caption{The distribution of the absolute magnification for the 62 multiple images included in our lens model.} 
    \label{fig:histo_ampli}
\end{figure}

Finally, in Fig.~\ref{fig:histo_ampli} we show the distribution of the absolute magnification values measured at the positions of all multiple images. The individual values and their errors are reported in Table~\ref{table:multimagesummary}. The magnifications are in the range $[5\div 360]$. Note that, as explained in \cite{Meneghetti_2017} the magnification estimates can be affected by significant systematic errors in the high-magnification regime. The reported errors only include the statistical uncertainties. 

\section{Summary and conclusions}
\label{sec:conclusions}

In this paper, we presented the strong lensing model of the galaxy cluster \CL at redshift $z=0.443$. This cluster is known mostly because it hosts the Sunburst arc, a very extended system of gravitational arcs, whose discovery was reported by \cite{dahle_2016}. These arcs are the highly magnified images of a source at redshift $z=2.3702$. Twelve multiple images of a single compact star forming region leaking Lyman continuum radiation were reported in these arcs \citep{RiveraT_2017,RiveraT_2019}. Using archival HST and MUSE data, we identified the multiple images of other twelve star forming knots in the Sunburst source. Moreover, we found three additional multiple image systems distributed over a circular region of radius $\sim 200$\arcsec\ around the cluster BCG. They span a redshift range between $z\sim 1.2$ and $z\sim 2.4$. In total, we secured 62 multiple images belonging to 17 families. 

We used the public parametric software \LT\ \citep{Kneib_lenstool,Jullo_Kneib_lenstool} to build a model of the cluster based on the detected multiple images. The model includes both large- and small-scale mass components, each described by dPIE lens models. The small-scale components consists of sub-halos associated with 197 cluster galaxies and with 3 additional galaxies along the line-of-sight. A sub-sample of 194 cluster galaxies are included in the model using scaling relations linking their central velocity dispersions and truncation radii to their luminosities in the F160W band. Following the method of \cite{Bergamini_2019} \citep[see also ][]{Bergamini_2020}, the $\sigma-L$ relation is calibrated using kinematic constrains derived from the analysis of the MUSE spectra of 15 cluster members. The remaining cluster galaxies and the 3 other galaxies along the line-of-sight are optimized individually because of their location is close to some of the multiple images. In particular, these galaxies perturb the Sunburst arc system. 

Our main results can be summarized as follows:
\begin{enumerate}
    \item the lens model reproduces accurately the positions of the multiple images used as constraints. The total r.m.s separation between model-predicted and observed image positions is $\Delta_{rms}\sim 0.14$\arcsec;
    \item the large-scale mass distribution of \CL\ is well described by the combination of two halos, one centered at a distance of few arcsec from the BCG and one centered in the north-eastern region of the cluster;
    \item the cluster galaxies follow scaling relations whose slopes are consistent with those previously published for other galaxy clusters at similar redshifts;
    \item the cluster mass distribution is characterized by a small ellipticity. The mass projected within $\sim 200$ kpc is $\sim 2\times 10^{14} \;M_\odot$. We find that the mass fraction in sub-halos within the same region is smaller than $10\%$;
    \item we used the lens model to compute maps of the magnification and of its uncertainty. The relative magnification errors are larger in the north-east region of the lens plane, where no multiple images are available to constrain the cluster mass model. These maps allowed us to measure the magnifications of all multiple images used to build the lens models. The magnification values are in the range $\mu\sim[5\div 360]$;
    \item the six perturbers near the upper segments of the Sunburst arc system create a complex network of resonant tangential caustics on the source plane at $z_s=2.3702$. In particular, there are regions in the source plane where several of these caustics overlap. The image multiplicity of some of the knots in the Sunburst arc is so high because the source intersects several of these regions.  
\end{enumerate}

The lens model we have produced will be used in several future works. First of all, it will allow to further characterize the physical properties of the source originating the Sunburst Arc (Vanzella et al., 2021). Second, it will be used to verify the hypothesis on the transient nature of the object recently reported by  \cite{Vanzella_bowen_2020}. In particular, we will use the time-delay maps derived from our model to predict the re-appearance of this source at different positions in the Sunburst arc. 

The major limitations of the model are due to the lack of constraints in some regions of the lens plane, in particular north-east or near the BCG. In order to improve the robustness of the model in these areas, deeper exposure times with both HST and MUSE will be helpful. The cluster has a large Einstein radius ($\sim 29$\arcsec\ for $z_s=2.3702$), thus we expect that many other systems of multiple images will emerge in the case of deeper exposures.  

The lens model of \CL\ presented here, together with the catalogs of cluster members and multiple images, will be publicly available for download from our server\footnote{ \href{http://www.fe.infn.it/astro/lensing/}{www.fe.infn.it/astro/lensing}}. Other studies report on the existence of a lens model of \CL\ \citep{2020MNRAS.491.4442L,RiveraT_2019}, which, however, is still un-published and therefore not publicly available (Sharon et al., in prep.). Thus, we defer to a future study the comparison of our reconstruction with other lens models. 
    
\begin{acknowledgements}
      This project is partially funded by PRIM-MIUR 2017WSCC32. PB and MM acknowledges financial support from ASI through the agreement ASI-INAF n. 2018-29-HH.0. CG acknowledges support by VILLUM FONDEN Young Investigator Programme through grant no. 10123. We acknowledge funding from the INAF ``main-stream'' grants 1.05.01.86.20 and  1.05.01.86.31. GBC thanks the Max Planck Society for support through the Max Planck
Research Group for S. H. Suyu and the academic support from the German
Centre for Cosmological Lensing."
\end{acknowledgements}

%
%

\bibliographystyle{aa}
\bibliography{bibliography}

\appendix
\section{Model refinement}
In this Section, we summarize the steps that led to the construction of the lens model presented in the paper. 

In the beginning, we fit only a small fraction of the lensing constraints. Through a series of iterations, we then refine the model by additional families of multiple images, starting with the best identified ones. As explained earlier, all mass components are modeled using dPIE lens model.

A crucial ingredient for the model optimization is the definition of the priors on the model parameters. Our initial guesses on several parameters (like, e.g., the central velocity dispersion, ellipticity, and orientation of the large-scale mass components or some Sunburst arc perturbers) are motivated by the geometry and angular separations of the multiple image systems. The priors on the parameters of the scaling relations used to describe the population of cluster members are based on the kinematic measurements, as discussed in Sect.~\ref{sec: galaxy_mass_distribution}. At each iteration, the priors are re-evaluated and changed, if necessary, based on the outcome of the modeling.

At all iterations, the lens model includes at least two large-scale potentials describing the smooth cluster dark-matter halo and its asymmetries, 194 galaxy-scale potentials associated to the cluster galaxies whose positions are not close to the multiple images, and six potentials associated to the Sunburst arc perturbers. 

\begin{figure*}
    \centering
    \includegraphics[width=0.9\linewidth]{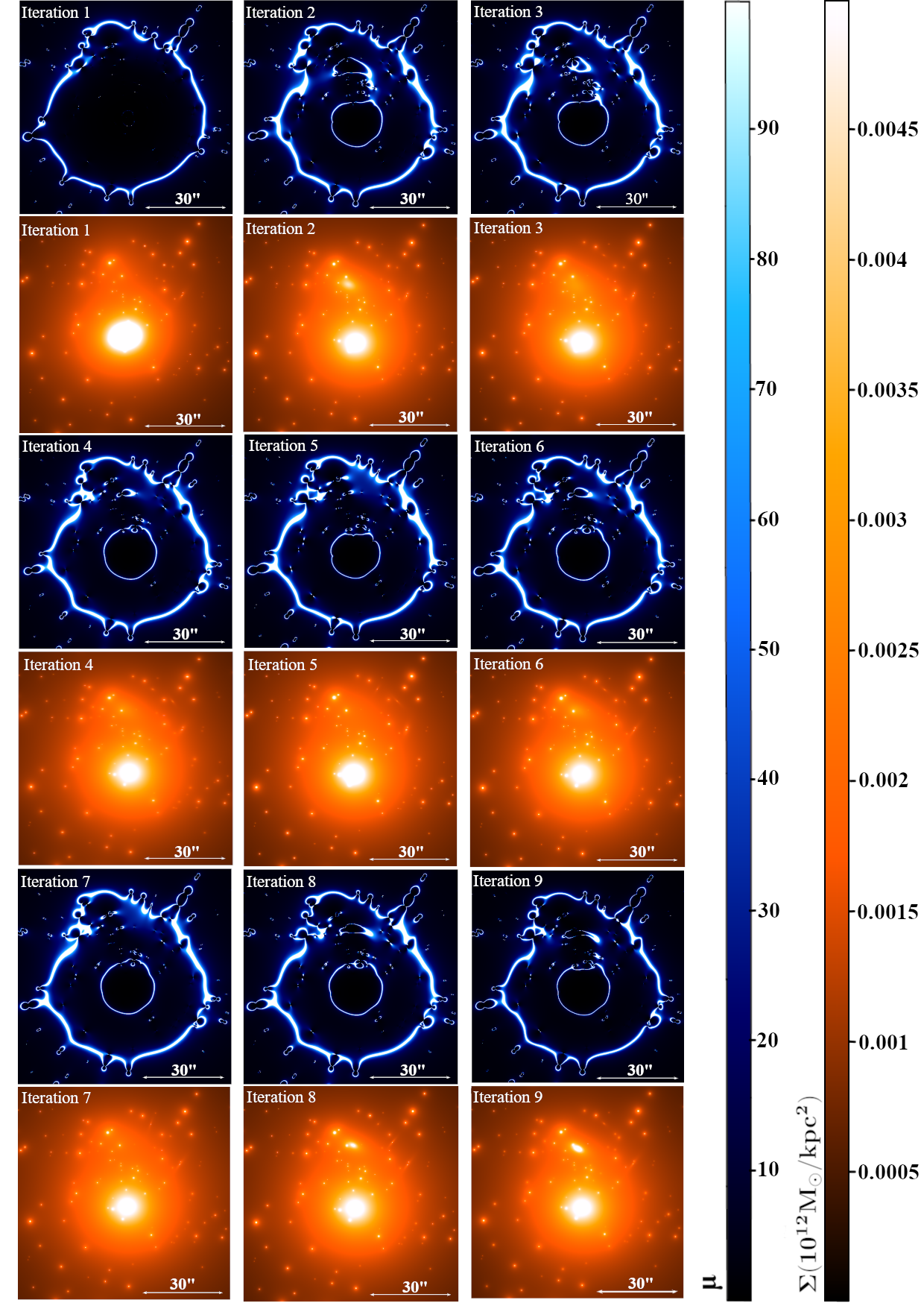}
    \caption{Surface density and magnification maps after each iteration of the model refinement. The magnification maps are visualized using a blue-gradient color scale, saturated at $\mu=100$. They are calculated assuming a source redshift of $z_s=2.3702$. The white color indicates the location of the lens critical lines for this source redshift.}
    \label{fig:modelev}
\end{figure*}

\begin{itemize}
\item \textbf{Iteration 1:} the starting model is based only on family 5.1.  The main cluster-scale halo is centered on the BCG. We vary its velocity dispersion  between 400 km $s^{-1}$ and 1200 km $s^{-1}$. The ellipticity is assumed to be small ($e<0.3$), since the Sunburst arc shape suggests that the cluster is quite circular. The core radius varies between $0''$ and $20''$. The second cluster-scale halo position is left free to vary in the north-eastern area from the BCG, near the first segment of the Sunburst arc, and its velocity dispersion is assumed to be in the range $250 \div 750$ km $s^{-1}$. In addition to the basic components listed above, in this iteration we use a potential to account for the presence of the BCG. In this case, we assume a circular mass distribution. This choice is motivated by the large distance between the BCG and the nearest strong lensing constraints. We assume that these constraints are insensitive to the BCG ellipticity. We fix the BCG core radius to be $0.01''$. We assume that its velocity dispersion is in the range $10-600$ km $s^{-1}$, and that the truncation radius is between $0.1$\arcsec\ and $20$\arcsec. In this first model, all the Sunburst arc perturbers are described by circular dPIEs. We set the priors on the velocity dispersion of each perturber based on its luminosity. Overall, we allow velocity dispersions in the range $50\div150$ km $s^{-1}$. For all of them, we fix the core radius to be $0.01$\arcsec, while we assume that the truncation radii are in the range $0.1\div10$\arcsec. The total r.m.s. separation $\Delta_{rms}$ achieved after optimizing this model is $0.31$\arcsec;

\item \textbf{Iteration 2:} at this stage we add Sys-3, Sys-4 and families 5.2 and 5.3 to the model constraints. Given the lack of constraints in the central region of the cluster, we choose to remove the potential describing the BCG, which is then incorporated in the main cluster-scale potential reducing the number of free parameters. We model the perturbers $Def-$N-ARC-R and $Def-1298$ using elliptical mass distributions. We assume that their ellipticities are in the ranges $0-0.7$ and $0.2-0.9$, respectively. Optimizing this model, we are able to reproduce well the positions of all multiple images, except image 5.1a, for which we find a large offset between observed and model-predicted positions. 
For this model, we obtain $\Delta_{rms}=0.12$\arcsec;

\item \textbf{Iteration 3:} we fit the model using also families 5.7 to 5.9. Given that the previous model worked well even without modeling the BCG independently for the large-scale components, we continue to follow this approach. In order to accommodate the new families of multiple images, we modify the priors on the velocity dispersion of the main cluster-scale component, which is now assumed to be in the range $700-1200$ km $s^{-1}$. We also assume that the second cluster-scale halo is free to move in DEC between $15$\arcsec\ and $31$\arcsec\ from the BCG position. Changing the position of the second cluster-scale component reduces significantly the offset of image 5.1a. After iteration 3, the total r.m.s separation is $\Delta_{rms}$ of $0.14$\arcsec;

\item \textbf{Iteration 4:} we include also families 5.4(a,b,c), 5.6 and 5.11. With respect to the last model, we modify the priors on the ellipticity of the main cluster-scale halo, which we assume to be in the range $0.0\div0.5$. We also change the priors on the ellipticity of $Def-$N-ARC-R, which is free to vary in the range $0.0\div0.9$. In addition, we model the perturber $Def-$BK-NW-L using an elliptical mass distribution, adding its ellipticity and position angle to the free parameters. For the ellipticity, we allow any value in the range $0-0.9$. The resulting model predicts the existence of additional multiple images of family 5.4, that are already identified as images 5.4(d,e,f). More multiple images of families 5.7, 5.8, 5.9 and 5.11 are also predicted, for instance in the third segment of the Sunburst arc, where the depth and resolution of the HST observations do not allow us a robust identification of these images.  The total r.m.s. separation for this model is $\Delta_{rms}=0.13$\arcsec;

\item \textbf{Iteration 5:} in this iteration, we use also the images 5.4(d,e,f), and families 5.12, 5.13(g,h). These images help to refine the model near the upper segments of the Sunburst arc. For example, the images of families 5.12 and 5.13 bracket the lens critical line close to perturbers $Def-$BK-N-NW and $Def-$BK-NW-L, as shown in inset $D$ of Fig.~\ref{fig:multiple} and in the purple-framed sub-panel of Fig.~\ref{fig:deflectors}. Unfortunately, the model fails to correctly reproduce images 5.1h, 5.1i, and 5.1l, and the passage of the critical line between these images, as shown in the yellow-framed sub-panel of Fig.~\ref{fig:deflectors}. The model produced in this iteration has total r.m.s. separation of  $\Delta_{rms}=0.17$\arcsec;

\item \textbf{Iteration 6:} to improve the accuracy of the lens model near the perturber $Def-1298$, we add family 5.14 as constraint in the next iteration. Indeed, images 5.14i and 5.14l are very close to images 5.1i and 5.1.h. Adding these constraints forces the cluster critical line to pass in between these images. We obtain a total r.m.s. separation of $\Delta_{rms}=0.19$\arcsec;

\item \textbf{Iteration 7:} at this stage, we use also family 5.5.  With the addition of this family the $\Delta_{rms}$ becomes smaller ($0.14$\arcsec), but the critical line passing between families 5.12 and 5.13 is now misplaced;

\item \textbf{Iteration 8:} to fix the issue emerged in iteration 7, we allow the center of $Def-$BK-NW-L to shift from its observed position by $4$\arcsec in both directions. Note that this perturber is in the cluster background ($z=0.7346$), thus its observed and true positions are expected to be different due to lensing. The model total r.m.s. is now $\Delta_{rms}=0.15$\arcsec.

\item \textbf{Iteration 9:} finally, we also include Sys-2. The redshifts of the two families are additional free parameters. With this adjustment, we obtain our final model of \CL, which is characterized by a total $\Delta_{rms}$ of $0.14$\arcsec.

\end{itemize}
To illustrate the evolution of the lens model through all these iterations, we show the corresponding surface density and magnification maps (assuming a source redshift $z_s=2.3702$) in Fig.~\ref{fig:modelev}. 

\begin{table*}[]
\resizebox{\textwidth}{!}{%
\def\arraystretch{1.9}
\centering
\begin{tabular}{ccccccccccc}
\cline{1-5} \cline{7-11}
\textbf{ID} & \textbf{RA}   & \textbf{Dec}  & \textbf{z}                & \textbf{$\mu$}                 & \textbf{} & \textbf{ID} & \textbf{RA}    & \textbf{Dec}  & \textbf{z} & \textbf{$\mu$}                  \\ \cline{1-5} \cline{7-11}
2.1a        & 15:50:17.10 & -78:11:42.3 & \textsuperscript{*}$2.197^{+0.026}_{-0.025}$ & $7.4_{-0.2}^{+0.2}$     &           & 5.3h        & 15:50:01.19  & -78:11:07.9 & 2.3702      & $97.5_{-19.0}^{+21.0}$   \\
2.1b        & 15:50:00.34 & -78:11:25.3 & \textsuperscript{*}$2.197^{+0.026}_{-0.025}$ & $9.1_{-0.4}^{+0.3}$       &           & 5.3m        & 15:49:58.57  & -78:11:27.3 & 2.3702      & $15.3_{-0.9}^{+1.1}$     \\ \cline{1-5}
2.2a        & 15:50:17.29 & -78:11:40.8 & \textsuperscript{*}$2.194^{+0.022}_{-0.021}$ & $7.8_{-0.2}^{+0.2}$     &           & 5.3n        & 15:50:15.24  & -78:11:46.9 & 2.3702      & $13.0_{-0.5}^{+0.5}$      \\ \cline{7-11}
2.2b        & 15:50:00.33 & -78:11:26.3 & \textsuperscript{*}$2.194^{+0.022}_{-0.021}$ & $7.6_{-0.3}^{+0.3}$     &           & 5.4a        & 15:50:07.32  & -78:10:57.3 & 2.3702      & $71.7_{-16.1}^{+26.1}$   \\ \cline{1-5}
3a          & 15:50:16.12 & -78:11:41.7 & 2.393                     & $16.1_{-0.8}^{+1.0}$    &           & 5.4b        & 15:50:06.23  & -78:10:58.0 & 2.3702      & $45.7_{-6.2}^{+6.8}$     \\
3b          & 15:49:58.86 & -78:11:33.6 & 2.393                     & $12.3_{-0.6}^{+0.7}$    &           & 5.4c        & 15:50:05.93  & -78:10:58.4 & 2.3702      & $33.1_{-5.5}^{+6.3}$     \\ \cline{1-5}
4a          & 15:50:14.23 & -78:11:38.2 & 1.186                     & $9.2_{-0.4}^{+0.5}$     &           & 5.4d        & 15:50:04.64  & -78:10:59.5 & 2.3702      & $14.1_{-1.4}^{+1.4}$     \\
4b          & 15:50:02.16 & -78:11:24.0 & 1.186                     & $69.1_{-13.0}^{+16.5}$  &           & 5.4e        & 15:50:04.36  & -78:10:59.9 & 2.3702      & $6.8_{-1.6}^{+2.1}$      \\ \cline{1-5}
5.1a        & 15:50:07.40 & -78:10:57.2 & 2.3702                     & $105.3_{-37.5}^{+66.0}$ &           & 5.4f        & 15:50:04.23  & -78:11:00.2 & 2.3702      & $14.6_{-1.4}^{+1.7}$     \\ \cline{7-11}
5.1b        & 15:50:06.14 & -78:10:58.1 & 2.3702                     & $74.9_{-15.4}^{+23.8}$   &           & 5.5a        & 15:50:06.88  & -78:10:57.4 & 2.3702      & $36.7_{-5.0}^{+5.5}$       \\
5.1c        & 15:50:05.98 & -78:10:58.3 & 2.3702                     & $62.5_{-15.2}^{+25.4}$  &           & 5.5b        & 15:50:06.52  & -78:10:57.6 & 2.3702      & $36.6_{-4.2}^{+4.6}$     \\
5.1d        & 15:50:04.59 & -78:10:59.6 & 2.3702                     & $14.2_{-1.4}^{+1.5}$    &           & 5.5c        & 15:50:05.75  & -78:10:58.5 & 2.3702      & $12.2_{-1.5}^{+1.2}$     \\
5.1e        & 15:50:04.39 & -78:10:59.9 & 2.3702                     & $5.8_{-1.4}^{+1.7}$     &           & 5.5d        & 15:50:04.81  & -78:10:59.2 & 2.3702      & $13.7_{-1.4}^{+1.4}$     \\ \cline{7-11}
5.1f        & 15:50:04.20 & -78:11:00.2 & 2.3702                     & $13.8_{-1.1}^{+1.4}$    &           & 5.6a        & 15:50:07.04  & -78:10:57.1 & 2.3702      & $29.2_{-4.0}^{+5.3}$     \\
5.1g        & 15:50:02.22 & -78:11:04.9 & 2.3702                     & $11.0_{-1.5}^{+1.6}$    &           & 5.6b        & 15:50:06.50  & -78:10:57.4 & 2.3702      & $27.2_{-2.7}^{+2.7}$     \\
5.1h        & 15:50:00.37 & -78:11:10.7 & 2.3702                     & $56.2_{-5.1}^{+6.3}$    &           & 5.6c        & 15:50:05.69  & -78:10:58.4 & 2.3702      & $9.9_{-1.3}^{+0.9}$      \\
5.1i        & 15:49:59.96 & -78:11:12.4 & 2.3702                     & $79.4_{-10.9}^{+14.4}$   &           & 5.6d        & 15:50:04.82, & -78:10:59.0 & 2.3702      & $13.2_{-1.3}^{+1.4}$     \\ \cline{7-11}
5.1l        & 15:49:59.75 & -78:11:13.6 & 2.3702                     & $79.2_{-8.8}^{+13.2}$   &           & 5.7c        & 15:50:05.68  & -78:10:58.8 & 2.3702      & $14.1_{-1.6}^{+1.5}$     \\
5.1m        & 15:49:58.54 & -78:11:26.9 & 2.3702                     & $14.7_{-0.8}^{+0.9}$    &           & 5.7d        & 15:50:04.97  & -78:10:59.3 & 2.3702      & $16.2_{-1.8}^{+1.7}$     \\ \cline{7-11}
5.1n        & 15:50:15.09 & -78:11:47.5 & 2.3702                     & $14.5_{-0.6}^{+0.8}$    &           & 5.8c        & 15:50:05.59  & -78:10:58.7 & 2.3702      & $13.9_{-1.6}^{+1.5}$     \\ \cline{1-5}
5.2a        & 15:50:06.75 & -78:10:57.5 & 2.3702                     & $95.0_{-24.6}^{+44.7}$   &           & 5.8d        & 15:50:05.05  & -78:10:59.1 & 2.3702      & $17.4_{-1.9}^{+1.8}$     \\ \cline{7-11}
5.2b        & 15:50:06.59 & -78:10:57.6 & 2.3702                     & $97.5_{-24.9}^{+44.3}$  &           & 5.9c        & 15:50:05.55  & -78:10:58.6 & 2.3702      & $12.7_{-1.5}^{+1.4}$      \\
5.2c        & 15:50:05.75 & -78:10:58.6 & 2.3702                     & $13.9_{-1.7}^{+1.3}$      &           & 5.9d        & 15:50:05.06  & -78:10:58.9 & 2.3702      & $17.0_{-1.9}^{+1.7}$     \\ \cline{7-11}
5.2d        & 15:50:04.84 & -78:10:59.3 & 2.3702                     & $14.2_{-1.5}^{+1.3}$    &           & 5.11c       & 15:50:05.47  & -78:10:58.7 & 2.3702      & $19.4_{-2.3}^{+2.2}$     \\
5.2g        & 15:50:02.12 & -78:11:05.2 & 2.3702                     & $19.2_{-2.6}^{+2.7}$     &           & 5.11d       & 15:50:05.17  & -78:10:58.9 & 2.3702      & $23.7_{-2.8}^{+2.5}$      \\ \cline{7-11}
5.2h        & 15:50:00.92 & -78:11:08.8 & 2.3702                     & $57.2_{-5.9}^{+6.6}$    &           & 5.12g       & 15:50:01.80  & -78:11:05.8 & 2.3702      & $93.8_{-12.4}^{+15.1}$     \\
5.2m        & 15:49:58.56 & -78:11:27.1 & 2.3702                     & $15.0_{-0.9}^{+1.0}$     &           & 5.12h       & 15:50:01.44  & -78:11:06.8 & 2.3702      & $148.6_{-21.8}^{+26.1}$  \\ \cline{7-11}
5.2n        & 15:50:15.18 & -78:11:47.1 & 2.3702                     & $13.4_{-0.6}^{+0.6}$     &           & 5.13g       & 15:50:01.79  & -78:11:06.4 & 2.3702      & $264.5_{-61.2}^{+96.9}$  \\ \cline{1-5}
5.3c        & 15:50:05.64 & -78:10:58.7 & 2.3702                     & $12.9_{-1.5}^{+1.2}$     &           & 5.13h       & 15:50:01.55  & -78:11:07.1 & 2.3702      & $358.5_{-82.7}^{+121.9}$ \\ \cline{7-11}
5.3d        & 15:50:04.98 & -78:10:59.2 & 2.3702                     & $15.1_{-1.6}^{+1.5}$    &           & 5.14i       & 15:50:00.00  & -78:11:12.6 & 2.3702      & $71.1_{-6.4}^{+7.6}$     \\
5.3g        & 15:50:02.00 & -78:11:05.6 & 2.3702                     & $41.4_{-8.7}^{+11.0}$   &           & 5.14l       & 15:49:59.80  & -78:11:13.7 & 2.3702      & $86.6_{-6.1}^{+7.6}$      \\ \cline{1-5} \cline{7-11}
\end{tabular}%
}
\smallskip
\caption{Coordinates, redshifts, and magnification estimates of the 62 multiple images used to build the lens model of \CL. The multiple images are grouped by families, separated by the horizontal black lines. The redshifts are measured from the MUSE spectra. The magnification values are estimated from the lens model. We report the median values obtained by sampling the model parameter posterior distribution functions to derive 100 realizations of the magnification maps. The quoted errors correspond to the 16-th and 84-th percentiles of the magnification distributions at each image position. \textsuperscript{(*)} The redshifts of Sys-2 are estimated with \LT\ (see \Sec\ref{sec:caseofsys2}).}
\label{table:multimagesummary}
\end{table*}

\end{document}